\documentclass[aps,pra,%
10pt,
final,%
notitlepage,%
oneside,%
twocolumn,
nobibnotes,%
nofootinbib,%
superscriptaddress,%
centertags,
floatfix]%
{revtex4-1}
\usepackage{graphicx}
\usepackage{amsmath,amsthm,amssymb,amscd}
\DeclareMathOperator{\Tr}{Tr}

\usepackage{color}
\usepackage{hyperref}
\newcommand{\OS}{{S_{\mathrm{o}}}}
\newcommand{\Os}{{S_{\mathrm{o}}}}
\begin{document}

\title{Thermal-difference states of light: quantum states of heralded photons}

\author{D.~B.~Horoshko}\email{Dmitri.Horoshko@univ-lille.fr}
\affiliation{Univ. Lille, CNRS, UMR 8523 - PhLAM - Physique des Lasers Atomes et Mol\'ecules, F-59000 Lille, France}
\affiliation{B.~I.~Stepanov Institute of Physics, NASB, Nezavisimosti Ave.~68, Minsk 220072 Belarus}%

\author{S.~De~Bi\`evre}
\affiliation{Univ. Lille, CNRS, UMR 8524 - Laboratoire Paul Painlev\'e, F-59000 Lille, France}

\author{G.~Patera}
\affiliation{Univ. Lille, CNRS, UMR 8523 - PhLAM - Physique des Lasers Atomes et Mol\'ecules, F-59000 Lille, France}

\author{M.~I.~Kolobov}
\affiliation{Univ. Lille, CNRS, UMR 8523 - PhLAM - Physique des Lasers Atomes et Mol\'ecules, F-59000 Lille, France}

\date{\today}

\begin{abstract}
We introduce the thermal-difference states (TDS), a three-parameter family of single-mode non-Gaussian bosonic states whose density operator is a weighted difference of two thermal states. We show that the states of ``heralded photons'' generated via parametric down-conversion (PDC) are precisely those among the TDS that are nonclassical, meaning they have a negative $P$-function.  The three parameters correspond in that context to the initial brightness of PDC and the transmittances, characterizing the linear loss in the signal and the idler channels. At low initial brightness and unit transmittances, the heralded photon state is known to be a single-photon state. We explore the influence of brightness and linear loss on the heralded state of the signal mode. In particular, we analyze the influence of the initial brightness and the loss on the state nonclassicality by computing several measures of nonclassicality, such as the negative volume of the Wigner function, the sum of quantum Fisher information for two quadratures, and the ordering sensitivity, introduced recently by us [Phys. Rev. Lett. \textbf{122}, 080402 (2019)]. We argue finally that the TDS provide new benchmark states for the analysis of a variety of properties of single-mode bosonic states.
\end{abstract}

\maketitle

\section{Introduction}

An optical field in a single-photon state \cite{Eisaman11} is an essentially quantum object, interesting for fundamental science and having numerous applications in quantum technologies like quantum cryptography \cite{Gisin02,Sangouard11}, linear optics quantum computation \cite{Knill01,Kok07}, boson sampling \cite{Tillmann13,Shchesnovich14,Quesada18}, as well as in detector calibration \cite{KlyshkoBook,Brida06} and radiometry \cite{Rodiek17}. Single photons can be obtained from single emitters, such as quantum dots \cite{Somaschi16,Ding16}, color centers \cite{Rodiek17,Fedotov12,Sipahigil14} or organic molecules \cite{Chu17,Rezai18}, and by all-optical methods either directly \cite{Kilin95,Mogilevtsev10} or conditionally by photon heralding technique \cite{Zeldovich69,Hong86,Grangier86,Lvovsky01,URen04,Fasel04,Neergaard-Nielsen07,Brida11,Fortsch13,Kaneda15,Joshi18,Ansari18}. The latter technique consists in generating a photon pair in two modes and detecting a photon of one (idler) mode, preparing thus the other (signal) mode in a single-photon state. In the ideal case, where exactly one photon pair is generated in a given time window and there is no loss, the conditional state of the signal field is the one-photon Fock state. However, in realistic experimental conditions two and more pairs can be simultaneously generated by the source and the light collection and detection are accompanied by loss and non-unit quantum efficiency of the photodetector. These factors lead to the appearance of multiphoton and vacuum components in the state of the signal field. Traditionally, these components are viewed as an undesirable ``contamination'' of the single-photon state of the signal mode, whose ``purity'' is determined by the intensity correlation function at zero delay \cite{Fasel04,Neergaard-Nielsen07}. We argue in this paper that, in fact, an interesting family of non-Gaussian nonclassical states is produced in such photon heralding experiments, including, in particular, truncated thermal states \cite{Lee95,Navarrete15} and photon-added thermal states \cite{Agarwal92,Kiesel08}, well studied in the past.

We consider parametric downconversion (PDC) as a source of photon pairs and find an explicit expression for the density operator of the signal mode conditioned by photon detection in the idler one. We take into account three physical parameters of this photon heralding scheme: the strength of nonlinear coupling of the PDC process, measured by the initial brightness $\xi$, the transmittance  $\eta$ of the idler channel, which includes the quantum efficiency of the photodetector, and the transmittance $\mu$ of the signal channel. Our goal is to understand how these parameters affect the nature of the signal mode state and in particular its nonclassicality. An almost single-photon state is obtained in the signal mode when $\mu=1$ and in the limit of small $\xi$, i.e. in a spontaneous PDC regime, corresponding to rather low rate of photon pair generation. Thus, a trade-off exists between the brightness and the quality of conditionally generated single photons. We are interested in including high-gain PDC in our consideration and exploring if a similar trade-off exists between the brightness and the nonclassicality. For that purpose we use several measures of nonclassicality, including a recently introduced one, the ordering sensitivity (OS) of a quantum state~\cite{Bievre19}. The main results of our approach have been reported on the QTech'2018 conference \cite{Horoshko19WoC}. The structure of the conditional state was independently found in Ref.~\cite{Quesada18} in the context of boson sampling with heralded photons.

The paper is organised as follows. We first show in Sec.~\ref{sec:Definition} that the conditional states of the signal mode belong to a larger three-parameter family of single-mode optical states, that we call ``thermal-difference states'' (TDS) because they are weighted differences of two thermal states. In Sec.~III we show that the family of TDS includes -- possibly as limiting cases -- not only the single-photon state, but also such well-known states as the photon-added and photon-subtracted thermal states, the truncated thermal state, the thermal state and the vacuum.   It turns out that the signal mode states obtained through the photon heralding technique are precisely those TDS that are nonclassical, meaning their Glauber-Sudarshan $P$-function is non-positive.  This is shown in Sec.~IV, where we calculate the Glauber-Sudarshan $P$-function and the Wigner function of the TDS, which are simply obtained as weighted differences of Gaussians, making the analysis of their positivity straightforward. In Sec.~V we then analyze quantitatively the sensitivity of the degree of nonclassicality of the TDS to changes in their parameter values by computing several measures of nonclassicality, such as the negative volume of the Wigner function, the sum of quantum Fisher information for two quadratures, and the ordering sensitivity. We show that the three parameter family of TDS introduced here constitutes an excellent testbed for the study of various properties of single-mode photon states. Indeed, they include a number of well-known such states, are non-Gaussian except in some limiting cases, have easily computable and non-singular quasi-probability distributions (notably the $P$-function) and explicit occupation numbers. Sec.~VI summarizes the results and concludes the paper.


\section{Definition of thermal-difference states and their generation in a heralding experiment
\label{sec:Definition}}

\begin{figure}[ht!]
\center{\includegraphics[width=\columnwidth]{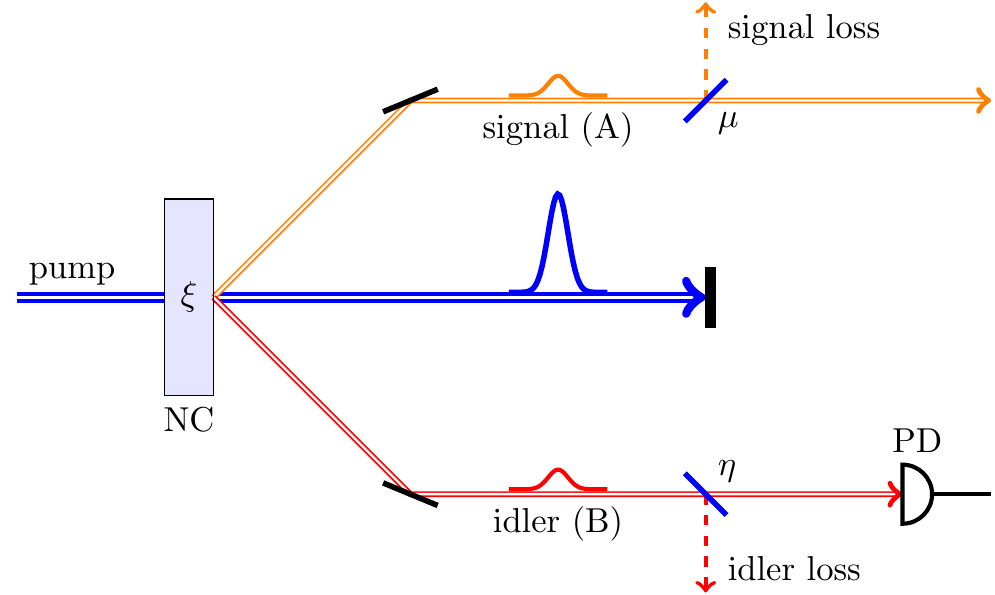}}
\caption{Schematic representation of the production of heralded photons by means of PDC. NC -- nonlinear crystal, where a two-mode squeezed state is generated with the rate $\xi$. PD -- photodetector, whose click conditions (heralds) a quasi-single-photon state in the mode A.}
\label{fig:setup}
\end{figure}

In the process of PDC an undepleted classical pump wave, passing through a nonlinear crystal, produces signal and idler waves, see Fig.~\ref{fig:setup}. In a typical PDC scenario these waves are multimode, and a set of Schmidt modes \cite{Law00,Horoshko12} can be defined for each wave such that a signal mode is correlated to the corresponding idler mode and the joint state of two photons is entangled. However, for the production of heralded photons a single-mode regime can be realized by tuning the pump pulse spectral width to that of the phase-matching function \cite{Grice01}. Let us accept that this regime is realised. We denote the signal mode by A and the idler mode by B and ascribe to these modes annihilation operators $a$ and $b$ respectively. The joint state of the two modes at the crystal output is \cite{BarnettBook}
\begin{eqnarray}\label{squeezing}
|\psi(r)\rangle_{AB} &=& e^{r(a^\dagger b^\dagger - ab)}|0\rangle_A|0\rangle_B \\\nonumber
&=& \mathrm{sech}\,r \sum_{n=0}^\infty \tanh^nr |n\rangle_A|n\rangle_B,
\end{eqnarray}
where $|n\rangle$ is the $n$-photon Fock state, and $r$ is the degree of squeezing determined by the pump amplitude, the nonlinear susceptibility of the crystal and its length.

The state of the signal mode alone can be obtained by tracing the above state over the space of the idler mode. As a result we obtain a thermal state \cite{Titulaer65}
\begin{equation}\label{rhotherm}
\rho_{\mathrm{th}}(\xi) = \left(1-\xi\right)\sum_{n=0}^\infty \xi^n|n\rangle\langle n|,
\end{equation}
where the parameter $\xi=\tanh^{2}r \in [0,1)$ is a parameter related to the temperature $T$ and the mean photon number $\langle n\rangle_{\mathrm{th}}$ by
\begin{equation}\label{q}
 \xi= e^{-\frac{\hbar\omega}{k_BT}} = \frac{\langle n\rangle_{\mathrm{th}}}{\langle n\rangle_{\mathrm{th}}+1},
\end{equation}
with $\omega$ being the circular frequency of the mode and $k_B$ the Boltzmann constant. As shown by Eq.~(\ref{rhotherm}), the thermal state is diagonal in the Fock basis and the number of photons follows the geometric distribution. The condition $\xi=0$ corresponds to zero temperature and zero mean photon number, i.e. to the vacuum state $\rho_{\mathrm{th}}(0)=|0\rangle\langle0|$. The opposite limit $\xi\to1$ corresponds to infinitely growing temperature. In general, $\xi$ is a monotonically increasing function of the temperature $T$ and can be considered as ``alternative temperature''. On the other hand, in the context of PDC, $\xi$ gives the probability of observing at least one photon in mode A, so it has the meaning of the photon pair generation rate.

We are interested in finding the state of the mode A under condition of a click of the detector monitoring the mode B in the general case, where the losses in the idler and the signal channels are characterized by their intensity transmittances $\eta$ and $\mu$ respectively.

We consider first the simplest case of no loss and unit quantum efficiency of the detector, $\eta=\mu=1$. A detector not resolving the number of photons and having unit quantum efficiency is characterized by the positive-operator valued measure (POVM) consisting of just two operators: operator $\Pi_{\text{off}}^B = |0\rangle_B{}_B\langle0|$ corresponding to no click and operator $\Pi_{\text{on}}^B = \mathbb{I}_B- |0\rangle_B{}_B\langle0|$ corresponding to a click. Here $\mathbb{I}_B$ is the identity operator for the mode B. Under the condition of observing a click at the detector monitoring the mode B, the conditional (unnormalized) state of the mode A is
\begin{eqnarray}\label{rhoA}
\tilde\rho^A &=& \mathrm{Tr}_B\left\{\Pi_{\text{on}}^B|\psi(r)\rangle_{AB}{}_{AB}\langle\psi(r)|\right\}\\\nonumber
&=& \mathrm{sech}^2\,r \sum_{n=1}^\infty \tanh^{2n}r |n\rangle_A{}_A\langle n|\\\nonumber
&=& \rho_{\mathrm{th}}(\xi) - (1-\xi)|0\rangle_A{}_A\langle 0|.
\end{eqnarray}
Upon normalisation of Eq.~(\ref{rhoA}) we obtain the conditional state of the mode A
\begin{equation}\label{rhoAnorm}
\rho^A = \frac{\tilde\rho^A}{\Tr\{\tilde\rho^A\}} = \frac{1-\xi}{\xi} \left(\frac{\rho_{\mathrm{th}}(\xi)}{1-\xi} - |0\rangle\langle 0|\right),
\end{equation}
which corresponds to the ``truncated thermal state'' \cite{Lee95} or the ``vacuum-removed thermal state'' \cite{Navarrete15}. This state contains no vacuum component, which has a simple physical explanation: in the absence of losses a click in mode B corresponds to the presence of at least one photon in the mode A. Note that in the limit $\xi\to0$, $\rho^A$ converges to the single-photon state:
\begin{equation}\label{eq:singlephoton}
\lim_{\xi\to0}\rho^A =|1\rangle\langle 1|.
\end{equation}
So for a small
degree of squeezing $r$, the state of the signal is close to the one-photon Fock state.

Now we consider a more complicated scenario, where the quantum efficiency of the idler channel $0<\eta \leq 1$ can be less than 1, but there are no losses in the signal channel. In this case the POVM of the detector is given by the operator \cite{Hogg14}
\begin{equation}\label{Pi}
\tilde\Pi_{\text{off}}^B = \sum_{n=0}^\infty (1-\eta)^n|n\rangle_B{}_B\langle n| = \frac{\rho_{\mathrm{th}}(1-\eta)}\eta,
\end{equation}
corresponding to no click, and the operator $\tilde\Pi_{\text{on}}^B = \mathbb{I}_B-\tilde\Pi_{\text{off}}^B$, corresponding to a click. Under condition of observing a click at the detector B, the conditional (unnormalized) state of the mode A reads
\begin{eqnarray}\label{rhoA2}
\tilde\rho^A &=& \mathrm{Tr}_B\left\{\tilde\Pi_{\text{on}}^B|\psi(r)\rangle_{AB}{}_{AB}\langle\psi(r)|\right\} \\\nonumber
&=& \rho_{\mathrm{th}}(\xi) - \frac{1-\xi}{1-\xi(1-\eta)}\rho_{\mathrm{th}}(\xi(1-\eta)),
\end{eqnarray}
which is a weighted difference of two thermal states of mode A.

The final scenario includes losses of the mode A, which are modelled by a beam splitter with the intensity transmittance $\mu \in (0,1]$, whose reflected field is traced out. The resulting channel $\rho\to\Phi(\rho)$ is known as the lossy quantum channel. When a field in a thermal state $\rho_{\mathrm{th}}(\xi)$ passes through such a channel, the state of the transmitted field is a thermal state with a lower temperature $\rho_{\mathrm{th}}(q)$. The value of $q$ can be found from the transformation of the mean photon number $\langle n\rangle=\mu\langle n\rangle_0$, where $\langle n\rangle_0=\xi/(1-\xi)$ and $\langle n\rangle=q/(1-q)$ are mean photon numbers at the input and the output of the beam splitter respectively. Solving this equation for $q$, we find
\begin{equation}\label{params}
q = \frac{\mu\langle n\rangle_0}{\mu\langle n\rangle_0+1} = \frac{\mu \xi}{1-\xi(1-\mu)}.
\end{equation}
The lossy quantum channel $\Phi(\rho)$ is linear in the input density operator $\rho$. Applying the transformation $\rho_{\mathrm{th}}(\xi)\to\rho_{\mathrm{th}}(q)$ to both summands of Eq.~(\ref{rhoA2}) and normalizing the resulting state, we arrive after some algebra at the state of the signal mode, which we call a ``thermal-difference state'' and which is given by the following expression:
\begin{eqnarray}\label{rho-}
\rho^{(-)}(q,p,{d}) &=&\mathcal{C} \left(\frac{\rho_{\mathrm{th}}(q)}{1-q} - {d}\frac{\rho_{\mathrm{th}}(qp)}{1-qp}\right) \\\nonumber
&=& \mathcal{C} \sum_{n=0}^\infty \left(q^n-{d}(qp)^n\right)|n\rangle\langle n|,
\end{eqnarray}
where
\begin{eqnarray}\label{C}
\mathcal{C} &=& \frac{\left(1-q\right)\left(1-qp\right)}{1-qp-{d}\left(1-q\right)}\\\nonumber
&=& \frac{1-\xi}{\eta \xi}\frac{1-\xi(1-\eta)}{1-\xi(1-\mu)} 
\end{eqnarray}
is a normalization factor, guaranteeing that $\Tr \rho^{(-)}(q,p,{d})=1$.  The  parameters $(q,p,{d})$ vary from $0$ to $1$ and are related to the physical parameters $\xi\in [0,1)$, $\eta, \mu\in(0,1]$  by Eq.~(\ref{params}) and
\begin{eqnarray}\label{params2}
{d} &=& \frac{1-\xi(1-\mu)}{1-\xi(1-\eta)(1-\mu)},\\\label{p}
p &=& (1-\eta){d}.
\end{eqnarray}
Note that, for the states produced in this manner through photon heralding with a non-zero detector quantum efficiency, $p$ is strictly less than ${d}$ due to Eq.~(\ref{p}). On the other hand, in Eq.~\eqref{rho-} the parameters $(q,p,{d})$ can in fact be allowed to vary in the full range from 0 to 1. More precisely, it is easy to see that for $q\in[0,1)$ and $p  \in [0, 1), {d}\in[0,1]$ the density operator, defined by Eq.~(\ref{rho-}), is always positive, since $\mathcal{C}>0$ and $q^n-{d}(qp)^n \ge0$ for any $n$, and normalizable. Hence Eq.~\eqref{rho-} defines a density operator for all these values, that fill thus a cube in the parameter space.  Each point of this cube corresponds to a positive density operator, i.e. some physical state of an optical mode.  The states obtained via photon heralding fill ``half'' of this cube, with parameters $0\leq p< {d}<1$. (See Fig.~\ref{fig:cube}.) In this half-cube the physical parameters $(\xi,\eta,\mu)$ are expressed through the mathematical ones $(q,p,{d})$ by the relations
\begin{eqnarray}\label{xi}
\xi &=& q+(1-q)\frac{1-d}{1-p},\\\label{eta}
\eta &=& 1-\frac{p}d,\\\label{mu}
\mu &=& \frac{q(d-p)}{(1-q)(1-d)+q(1-p)},
\end{eqnarray}
which are obtained by reversing Eqs.(\ref{params}),(\ref{params2}) and (\ref{p}).

It should be noted that the influence of losses on the photon statistics in the signal channel was studied before in detail \cite{Laurat04,DAuria12,Quesada15Thesis,Tiedau19}. However, the density operator of the conditional state, Eq.~(\ref{rho-}), was not written explicitly in these works.

It is often convenient to use the mathematical parameters $(q,p,{d})$ to describe the TDS because the simplicity of the expression in~\eqref{rho-} implies that a variety of quantities associated  to the TDS -- and in particular their quasi-probability distributions --  can be easily expressed in terms of those parameters. We will see several examples of this observation below. For their physical interpretation in terms of photon heralding, we will each time come back to the physical parameters $(\xi,\eta,\mu)$. If different methods for generating the TDS experimentally are found, one may expect different physical parameters to be relevant and perhaps different regions of the parameter cube to be realized experimentally.

In conclusion to this section, we have established that  the states generated in a photon-heralding experiment realize  one half of the parameter volume of the TDS, determined by operator positivity. The physical meaning of this part of the parametric space will become clear in Sec.~IV, where it will be shown that  it corresponds precisely to those TDS that are nonclassical.

\section{Exploring the family of thermal-difference states \label{sec:Family}}
\subsection{Parameter cube}

\begin{figure}[ht!]
\center{\includegraphics[width=0.7\columnwidth]{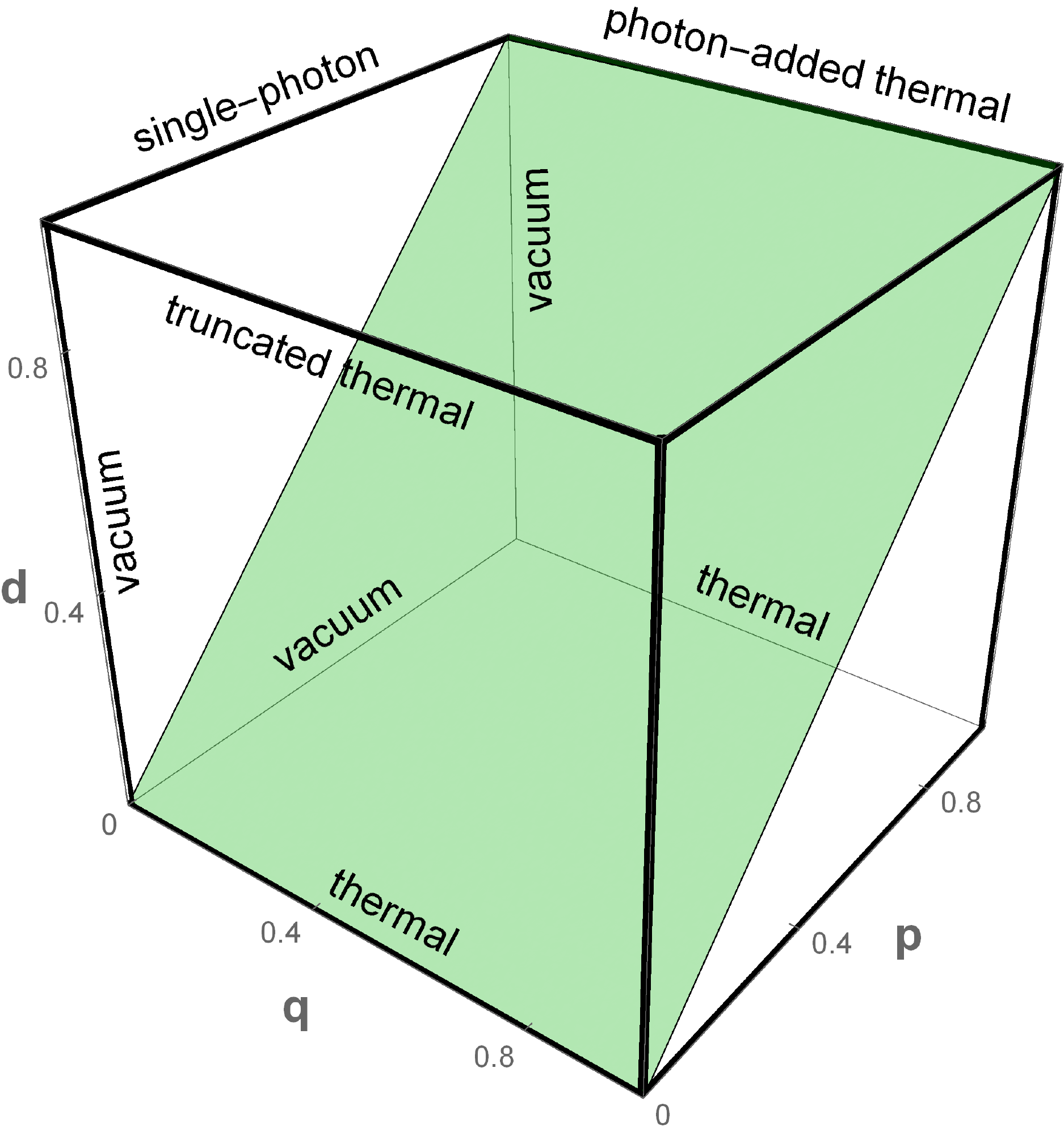}}
\caption{Parameter space of the thermal-difference states. Each point inside the cube corresponds to a positive density operator. The green plane ${d}=p$ is the border, above which the state corresponds to some value of the  physical parameters $(\xi,\eta,\mu)$ of Fig.~\ref{fig:setup}. The edge denominated ``single-photon'' is multi-valued and corresponds to a state $(1-\mu)|0\rangle\langle0| + \mu |1\rangle\langle1|$, where $\mu$ determines the angle at which the edge is approached. The edge denominated ``photon-added thermal'' is also multi-valued, see Sec.~\ref{sec:PATS}.}
\label{fig:cube}
\end{figure}

In this section we explore TDS as defined in Eq.~(\ref{rho-}) for the set of parameters $q\in [0,1), p\in [0, 1], {d}\in [0, 1]$ and study their physical meaning in the context of a photon heralding experiment. The density operator of these states is a weighted difference of two thermal states $\rho_{\mathrm{th}}(q)$ and $\rho_{\mathrm{th}}(qp)$, where $p\in[0,1]$, i.e. the temperature of the second state is lower than or equal to that of the first state. The parameters $(q, p, {d})$ constitute a cube of which the face $q=1$ has been removed, see Fig.~\ref{fig:cube}.

The points in the face $q=1$ of the parameter cube have no particular physical meaning.  Indeed, the limit in which $q\to1$ at fixed $p,{d}$,  corresponds to infinitely growing temperature, and the limiting value $q=1$ does not correspond to a density operator.

To understand the meaning of the points on the other faces, we consider below the limiting procedure in detail.

\subsection{Single-photon state \label{sec:SPS}}

On the face $q=0$, the following limiting states can be identified. When ${d}<1$, the limit $q\to0$ yields the vacuum state:
\begin{eqnarray}\label{qto01}
0\leq {d}<1, 0\leq p\leq1 \Rightarrow \lim_{q\to0}\rho^{(-)}(q,p,{d}) &=& |0\rangle\langle0|.
\end{eqnarray} Note that this is true also if $p<{d}$, which means that the states $\rho^{(-)}(q,p,{d})$, nonclassical under that assumption, approach the vacuum state, which is classical. The physical conditions corresponding to this limit become clear from Eqs.~(\ref{xi}) and (\ref{mu}), which give $\xi\to(1-d)/(1-p)$, and $\mu\to0$. Thus, the vacuum state is a result of increasing loss in the signal channel.

On the other hand, when ${d}=1$, and $p<1$, the limit $q\to0$ yields the one-photon Fock state which is the ultimate goal of the photon heralding technique, being a highly nonclassical state:
\begin{eqnarray}\label{qto02}
{d}=1, 0\leq p<1 \Rightarrow\lim_{q\to0}\rho^{(-)}(q,p,1) &=& |1\rangle\langle1|.
\end{eqnarray}
In terms of the physical parameters controlling the photon heralding technique, the regime ${d}=1$, $q\to 0$, $p<1$ corresponds to the situation where $\xi\to0$ and the signal transmittance $\mu$ takes on its maximal value $1$. Several considerations are important here. First, we see that the set of TDS is discontinuous on the edge $d=1$, $q=0$: the rest of the face $q=0$ corresponds to the vacuum, as shown by Eq.~(\ref{qto01}). Second, the limit, given by Eq.~(\ref{qto02}), is independent of $\eta$. Different values of $\eta$ correspond to different values of $p=1-\eta$, and thus to different points on the edge, all being single-photon states. Third, this regime requires $\mu=1$, which is hardly reachable in practice, and the realistic signal states, generated at $d<1$, are mixtures of the vacuum, single-photon and multiphoton components.

A typical regime of photon-heralding experiments consists in a very low $\xi$ at fixed $\eta,\mu<1$. In this regime we obtain
\begin{eqnarray}\label{xito0}
\lim_{\xi\to0}\rho^{(-)}(q,p,{d}) &=& (1-\mu)|0\rangle\langle0| + \mu |1\rangle\langle1|
\end{eqnarray}
independently of $\eta$. Geometrically this limiting state corresponds to the same edge $q=0$, $d=1$, which is approached at a different angle, determined by $\mu$. Equation~(\ref{xito0}) reduces to Eq.~(\ref{qto01}) or Eq.~(\ref{qto02}) in the limiting cases of total loss ($\mu=0$) or no loss ($\mu=1$) respectively.

A photon heralding experiment aims at generating a single-photon state and is typically characterized by two parameters: the brightness and the single-photon purity. The (per excitation) brightness is the probability of observing a coincidence at the detectors monitoring the signal and the idler modes. Accepting that the quantum efficiency of the detector monitoring the signal mode is included into $\mu$, we can obtain the brightness as the product of the probability of a click of the idler detector and the probability of having a nonzero photon number in the signal state, conditioned by this click:
\begin{eqnarray}\label{brightness}
\mathcal{B} &=& \left(1-\langle0|\rho^{(-)}(q,p,{d})|0\rangle\right)\Tr\tilde\rho^A \\\nonumber
&=& \left[1-\mathcal{C}(1-d)\right]\frac{\xi\eta}{1-\xi(1-\eta)}\\\nonumber
&=&  \xi\eta\mu\frac{1-\xi^2\bar\eta\bar\mu}{(1-\xi\bar\mu)(1-\xi\bar\eta)(1-\xi\bar\eta\bar\mu)},
\end{eqnarray}
where $\bar\eta=1-\eta$, $\bar\mu=1-\mu$. In the absence of losses, $\eta=\mu=d=1$, we find easily $\mathcal{B}=\xi$, i.e. $\xi$ can be understood as the initial brightness of PDC, before losses occur.

The single-photon purity is determined by the intensity autocorrelation function at zero delay calculated with the conditional signal state, Eq.~(\ref{rho-}):
\begin{eqnarray}\nonumber
g^{(2)}(0) &=& \frac{\langle a^{\dagger 2}a^2\rangle}{\langle a^\dagger a\rangle^2}
= 2\frac{1-df(1+p^2f^2)+d^2p^2f^4}{1-2dpf^2+d^2p^2f^4}\\\label{g2}
&=& 1-\frac{(1-\xi\bar\eta)(1-\xi)}{(1-\xi^2\bar\eta)^2},
\end{eqnarray}
where $f=(1-q)/(1-qp)$. It is easy to see that, since $1+p^2f^2\ge2pf$, we have $g^{(2)}(0)\le2$. The maximal value of 2, corresponding to a thermal statistics, requires $pf=1$, implying $p=1$ or $\eta\to0$. As expected, in this limit the TDS approaches a thermal state, see below Sec.~\ref{sec:PATS}. However, this state cannot be generated in a photon-heralding experiment, where $p<d$. Moreover, as follows from Eq.~(\ref{g2}), for the state generated in this experiment the value of $g^{(2)}(0)$ does not depend on the signal transmittance $\mu$ and is always less than 1, which corresponds to photon antibunching. For a mixture of vacuum and the single-photon state $g^{(2)}(0)=0$, which means the highest single-photon purity. The brightness and the purity together determine how close the generated state is to the single-photon state: the brightness determines the fraction of vacuum, while the purity determines the fraction of 2 and more photons.

It is generally known that in photon heralding a high purity is possible only in exchange to a rather low brightness \cite{URen04,Fasel04,Neergaard-Nielsen07,Brida11,Fortsch13,Kaneda15,Joshi18,Ansari18}. For example, to have $g^{(2)}(0)=0.2$ in an experiment with realistic values $\eta=\mu=0.8$, one needs the initial brightness $\xi=0.0082$, which means that most signal pulses are empty. In our approach we are interested in nonclassical properties of the conditional state, Eq.~(\ref{rho-}). To this end, we characterize in Sec.~\ref{sec:nonclassicality} below the signal state by a nonclassicality measure rather than by  its single-photon purity, and establish the existence of a similar nonclassicality-brightness trade-off.

\subsection{Truncated thermal state \label{sec:TTS}}

Let us consider now the face $p=0$, $0<q<1$. In that case, Eq.~(\ref{rho-}) yields a TDS obtained simply by substracting part of the vacuum component from the thermal state $\rho_{\text{th}}(q)$. In particular, the edge $0<q<1$, $p=0,\,{d}=1$ corresponds to the truncated thermal state \cite{Lee95}, also called the vacuum-removed thermal state \cite{Navarrete15}, already encountered in Eq.~\eqref{rhoAnorm}.  This is as expected since, expressed in the physical parameters of a photon heralding experiment, $p=0$ and ${d}=1$  correspond to $\mu=1=\eta$, i.e. no loss. On the rest of the face $0<q<1$, $p=0$, $0<{d}<1$ the vacuum is partially removed, having the weight $\mathcal{C}(1-d)$. This situation corresponds to losses in the signal mode, but not in the idler mode. The resulting state can be called ``partially vacuum-removed state''. At the lower edge $0<q<1$, $p=0$, $d=0$ the vacuum is not removed at all and the corresponding state is just a thermal state. The weight of vacuum $\mathcal{C}(1-d)$ tends to 1 as $q\to0$ for any $d<1$, so that the ``partially vacuum-removed state'' tends continuously to the vacuum. For $d=1$ the vacuum component is absent and the limiting state at $q\to0$ is the single-photon state.

\subsection{Photon-added thermal state \label{sec:PATS}}

We now study the face $p=1$, $0<q<1$ of the cube. When ${d}<1$, one can compute Eq.~(\ref{rho-}) with $p=1$ to find  the thermal state $\rho_{\mathrm{th}}(q)$. If ${d}=1$, the expression (\ref{rho-}) is singular for $p=1$: in that case, the limit $p\to1$ yields
\begin{eqnarray}\label{rto1}
\lim_{p\to1}\rho^{(-)}(q,p,1) &=& \left(1-q\right)^2\sum_{n=1}^\infty nq^{n-1}|n\rangle\langle n| \\\nonumber
&=& (1-q)a^\dagger\rho_{\mathrm{th}}(q)a,
\end{eqnarray}
in which we recognize the density operator of the photon-added thermal state \cite{Agarwal92}, which is nonclassical and has been much studied, also experimentally \cite{Kiesel08}. At the limit $q\to0$ this state approaches continuously the single-photon state at the vertex of the parameter cube.

From Eq.~(\ref{p}) we see that in an experiment the regime $p=1$, $d=1$ can be approached only at a very low $\eta$. It means that the final brightness will be much lower than the initial one. However, the initial brightness can be made rather high in this regime, in contrast to the regime of single photon generation. For example, setting $\xi=0.5$, $\eta=0.01$, $\mu=0.8$, we obtain $p\approx d=0.99$, which means that the generated state is close to a photon-added thermal state with the temperature $q=0.44$. However, its brightness is $\mathcal{B}=0.009$, which is rather low. Again we meet a trade-off between the brightness and the quality of the generated state. Note that generation of an almost photon-added thermal state in the proposed scheme is simpler than that of Ref.~\cite{Kiesel08}, since it does not require a thermal seed for the signal field.

The family of TDS is not continuous on the edge~$p=1, {d}=1$. Again, as in Sec.~\ref{sec:SPS}, we have a multi-valued edge, where the state depends on the angle $\theta$ to the horizontal plane, at which the edge is approached. In particular, setting
$d=1-(1-p)\tan\theta$ and taking the limit $p\to1$, one finds, for all $0<q<1$,
\begin{eqnarray}\label{rto2}
\lim_{p\to1}\rho^{(-)}(q,p,d) &=& \frac{1-q}{q+(1-q)\tan\theta}\\\nonumber
&\times& \left( qa^\dagger\rho_{\textrm{th}}(q)a + \rho_\textrm{th}(q)\tan\theta\right),
\end{eqnarray}
which gives Eq.~(\ref{rto1}) at $\theta=0$ and a thermal state at $\theta\to\pi/2$. At $\theta=\pi/4$ we obtain the so-called photon-removed thermal state, $a\rho_{\textrm{th}}(q)a^\dagger$, which is classical \cite{Zavatta08}. Its generation in the considered experiment is impossible, because due to Eq.~(\ref{p}) we have $p<d$ for any non-zero transmittance of the idler channel.

We have therefore established that the face $p=1$ corresponds to thermal states, with a $q$-dependent temperature, except for the ${d}=1=p$ edge, the points of which can be associated to a variety of states, among which the photon-added or -removed thermal states. We have seen that the first can be realized in principle with the photon heralding technique provided that the transmittance of the signal channel is high and that of the idler channel low.

\subsection{Summary}

The full physically relevant parameter space for the TDS forms a cube with only one missing face, corresponding to $q=1$, see Fig.~\ref{fig:cube}. The face $q=1$ and its four edges correspond to infinite temperature and are not interesting from the physical point of view. The remaining eight edges of the parametric cube on the other hand correspond to several well-known states of a single-mode field: the vacuum, thermal states, truncated thermal states, photon-added and -removed thermal states as well as the single-photon state and mixtures of the vacuum with the single-photon state. They are marked in Fig.~\ref{fig:cube}. The rest of the cube contains the new TDS we introduced here.

The states lying above the plane $d=p$ in this three-parameter family can in principle all be obtained through the photon heralding technique. In the next section we show that these states are nonclassical, except for some limiting points.

\section{Quasiprobability distributions for thermal-difference states\label{sec:quasi}}
\subsection{General $s$-parameterized distribution}

A remarkable and useful property of the TDS is the simple structure of their quasiprobability distributions. A quantum state of a single-mode optical field with a density operator $\rho$ is fully characterized by its $s$-ordered quasiprobability distribution \cite{Cahill69b}:
\begin{equation}\label{W}
 W(\alpha,s) = \int\Tr\left\{\rho e^{\lambda \left(a^\dagger-\alpha^*\right)-\lambda^*\left(a-\alpha\right)+s|\lambda|^2/2}\right\} \frac{d^2\lambda}{\pi^2},
\end{equation}
where $s\in[-1,1]$ is a real parameter, taking values $1$, $0$ and $-1$ for the Glauber-Sudarshan $P$-representation, the Wigner representation, and the Husimi $Q$-representation respectively.

For a thermal state, given by Eq.~(\ref{rhotherm}), this function is a two-dimensional Gaussian \cite{Cahill69b}
\begin{equation}\label{Wtherm}
 W_{\mathrm{th}}(\alpha,s|\xi) = \frac{\kappa(\xi,s)}\pi e^{-\kappa(\xi,s)|\alpha|^2},
\end{equation}
with the inverse variance function defined as
\begin{equation}\label{G}
 \kappa(\xi,s) = \frac2{2\xi/(1-\xi)+1-s}.
\end{equation}
For a TDS we obtain from Eqs.~(\ref{rho-}) and (\ref{W}) a difference of two two-dimensional Gaussians:
\begin{eqnarray}\label{W-}
 &&W^{(-)}(\alpha,s|q,p,{d}) \\\nonumber
 &&= \mathcal{C} \left(\frac{W_{\mathrm{th}}(\alpha,s|q)}{1-q} -{d}\frac{W_{\mathrm{th}}(\alpha,s|qp)}{1-qp}\right).
\end{eqnarray}

\subsection{$P$-function}
\begin{figure}[ht!]
\centering
\includegraphics[width=0.8\columnwidth]{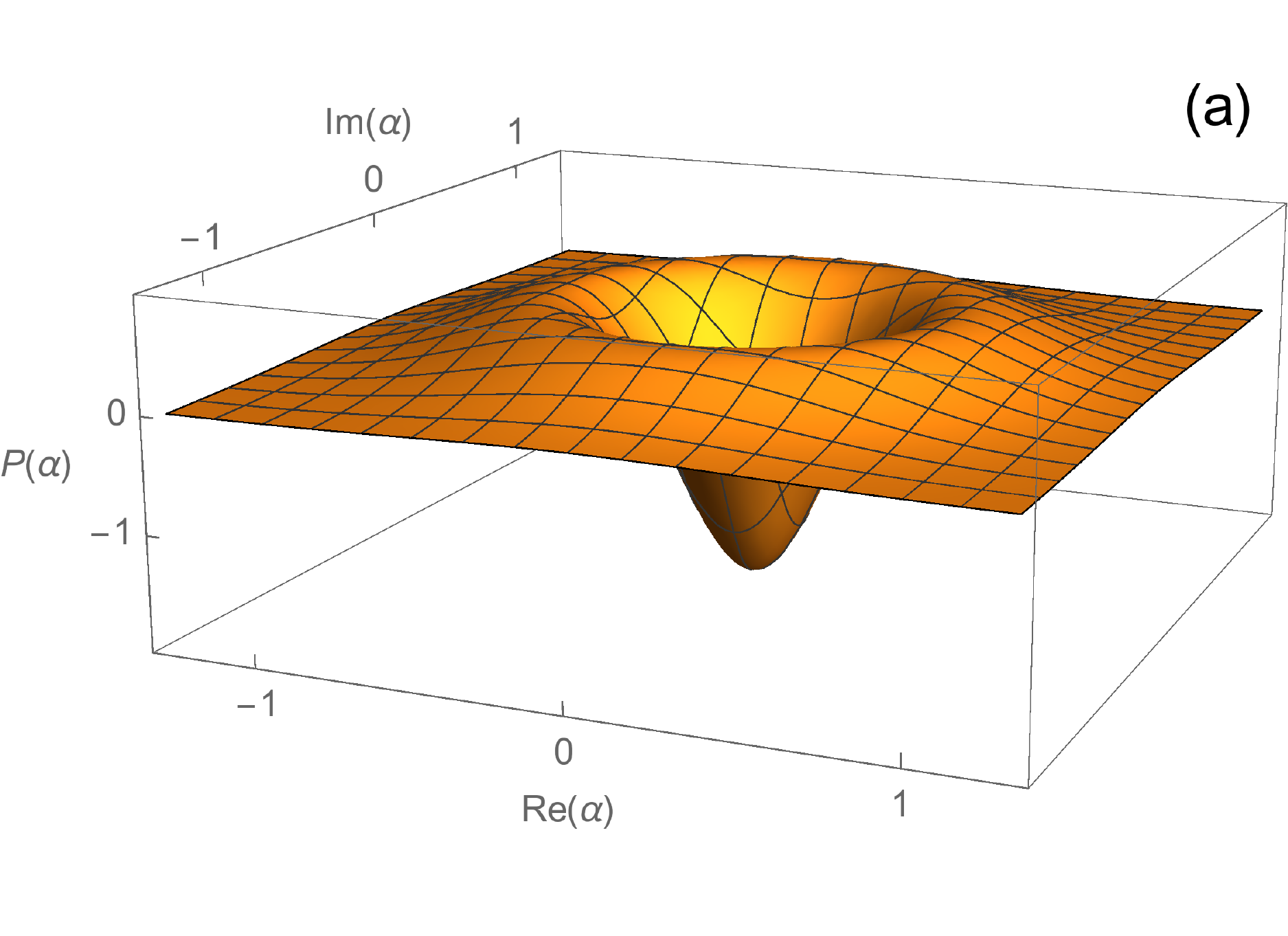}
\includegraphics[width=0.8\columnwidth]{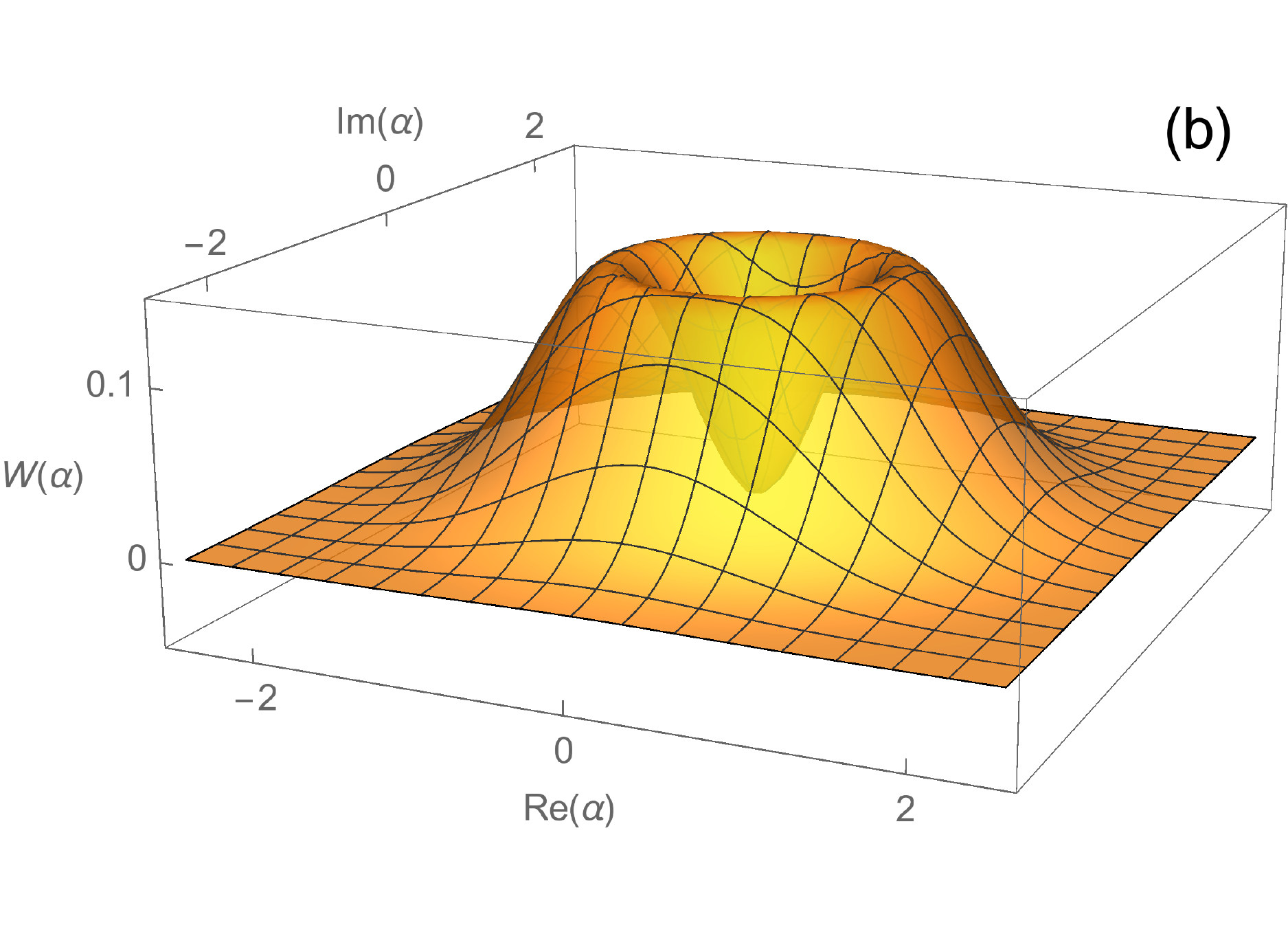}
\includegraphics[width=0.8\columnwidth]{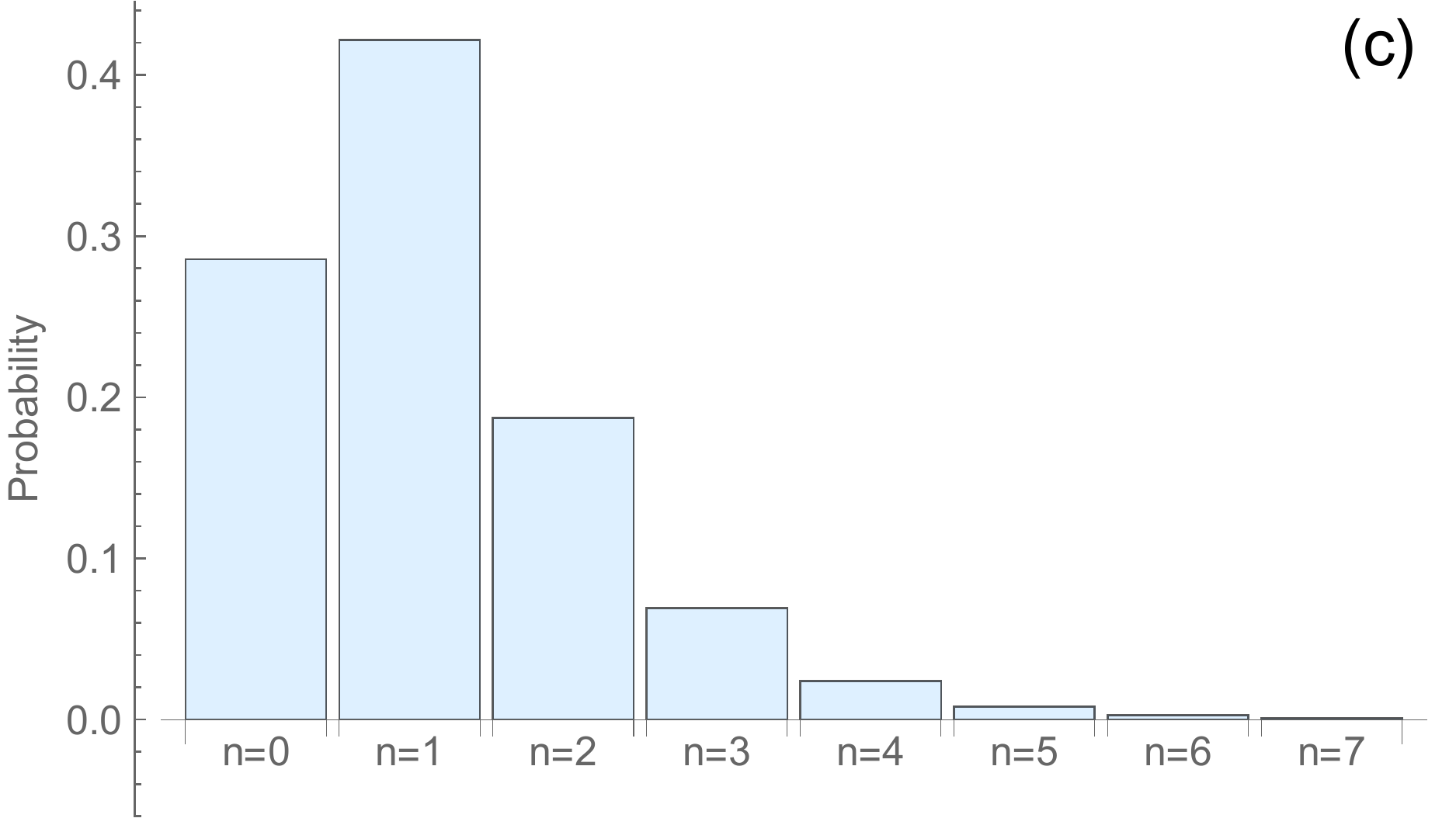}
\caption{(a) $P$-function, (b) Wigner function, and (c) photon number distribution for the TDS generated in a photon heralding experiment with $\xi=\eta=\mu=0.5$. This corresponds to the parameters $(q,p,{d})=(0.33,0.43,0.86)$. The negativity of the $P$-function  establishes the nonclassicality of this conditionally prepared state. Note that the Wigner function is positive for this set of parameters. The vacuum component ($n=0$) appears in the photon number distribution due to loss in the signal channel ($\mu<1$). \label{fig:P-}}
\end{figure}

The quasiprobability function takes a rather simple form at $s=1$, corresponding to the $P$-representation. Thus, defining $P^{(-)}(\alpha|q,p,{d}) = W^{(-)}(\alpha,1|q,p,{d})$, we obtain
\begin{eqnarray}\label{P-}
 &&P^{(-)}(\alpha|q,p,{d}) \\\nonumber
 &&= \frac{\mathcal{C}}{q\pi} \left(e^{-|\alpha|^2(1-q)/q} - \frac{d}{p}e^{-|\alpha|^2(1-qp)/qp}\right).
\end{eqnarray}
Note that this $P$-function is regular, without any singularities. As such, the TDS belong to a larger class of ``punctured'' states~\cite{Damanet18}. This function is shown in Fig.~\ref{fig:P-} together with the corresponding Wigner function and the photon number distribution $\langle n|\rho^{(-)}|n\rangle$ for experimentally feasible values of the parameters.

We can now easily identify among the TDS those that are nonclassical. We recall that a state $\rho$ of one mode is said to be classical if its $P$ function is positive everywhere. Otherwise it is said to be nonclassical.  For sufficiently low values of ${d}$ the second term in (\ref{P-}) is negligible with respect to the first one and the TDS approaches a thermal state,  known to be classical. Thus, at fixed values for $0<q<1$ and $0\leq p<1$, the nonclassical TDS, which are the most interesting ones from the point of view of quantum information, are the states with sufficiently high values of ${d}$. More precisely, when $P^{(-)}(\alpha|q,p,{d})$ is not everywhere positive its minimal value is reached at the origin, $\alpha=0$, and this value is
\begin{equation}\label{P-0}
 P^{(-)}(0|q,p,{d}) = \frac{\mathcal{C}}{q\pi} \left(1 -\frac{d}{p}\right).
\end{equation}
Since this is negative if and only if ${d}>p$ (and $q\ne0$), it follows that the TDS are non-classical iff ${d}>p$.   Note that due to (\ref{params2})  this is always the case for heralded photons if $\eta>0$. Thus, the half of the parametric cube which is realized in a photon heralding experiment is exactly the one corresponding to nonclassical TDS. Here we excluded the trivial case $q=0$, $d<1$, where the signal state is vacuum. For $d=1$, $q\to0$, corresponding to the single-photon state, the $P$-function is singular at the origin.

We note that the $P$-function of an optical mode, even when it takes negative values, can  in principle be reconstructed from a series of measurements. For example, the negative $P$-function of a photon-added thermal state \cite{Agarwal92} has been successfully reconstructed from experimental data \cite{Kiesel08}. As we have seen in the previous section, the photon-added thermal state is a particular limiting member of the family of TDS, generated for particular settings of the photon heralding experiment. A regular $P$-function can in principle be obtained from the experimental data in a way analogous to that of~\cite{Kiesel08} for other settings of such an experiment, except for extremely low values of the parameters $q$ and $p$. At low $p$ the $P$-function approaches a singular function
\begin{equation}\label{P-TTS}
 \lim_{p\to0}P^{(-)}(\alpha|q,p,{d})
 = \mathcal{C} \left[\frac{1}{q\pi}e^{-|\alpha|^2(1-q)/q} - d\delta(\alpha)\right]
\end{equation}
and cannot be reconstructed. At low $q$ the first term in Eq.~(\ref{P-TTS}) also approaches a delta-function.  In all other cases, it is in principle possible to experimentally establish the nonclassical nature of the generated state by reconstructing its $P$-function. Note that low $p$ corresponds to $\eta$ close to 1, which is hardly reachable in experiment, since the quantum efficiency is limited for modern single-photon detectors.

\subsection{Wigner function}

In contrast to the $P$-function, the Wigner function is not always negative for TDS generated in a photon-heralding experiment. Substituting $s=0$ and $\alpha=0$ into Eq.~(\ref{W-}) we obtain the value of the Wigner function at the origin
\begin{eqnarray}\label{W-00}
 &&W^{(-)}(0,0|q,p,d) = \frac{2\mathcal{C}}\pi \left(\frac{1}{1+q} -\frac{d}{1+qp}\right),
\end{eqnarray}
which is negative for $d>p+(1-p)/(1+q)$. Obviously, the latter condition is more restrictive than the condition of the $P$-function negativity, $d>p$. It is interesting to note, that the Wigner function of the signal mode in a photon-heralding experiment was successfully reconstructed from the experimental data \cite{Neergaard-Nielsen07} and its negativity was used as a signature of the state nonclassicality. Our analysis clearly shows that even for a positive Wigner function the state of the signal mode is nonclassical (cf. Fig.~\ref{fig:P-}), and its nonclassicality can be witnessed by reconstructing the $P$-function of the field. In general, the Wigner function is smoother than the $P$-function and, as a result, does not show the same sharp negative peak at low values of $q$ and $p$, as the $P$-function.

\subsection{Summary}

The regularity of their $P$-function, the simple explicit form of their quasi-probability distributions and occupation numbers, together with the simplicity of the nonclassicality condition, ${d}>p$, ensure that the TDS constitute an excellent benchmark to test the efficiency of  various nonclassicality witnesses and measures.  A number of pure quantum states of a single-mode optical field are traditionally used for this purpose, the most popular being the Fock state, the squeezed state \cite{Kolobov99} and the Schr\"odinger cat state, either with two \cite{Yurke86,Horoshko98} or multiple components \cite{HarocheBook,Horoshko16}.  Mixed states  typically used to that end include squeezed thermal states \cite{Kim89}, photon-added thermal states \cite{Agarwal92} and thermalized cat states \cite{Lee11}. The new family of TDS we introduce here widens this class significantly. In the next section we show that they allow one to analyze analytically a number of nonclassicality witnesses and measures.


\section{Nonclassicality of thermal-difference states \label{sec:nonclassicality}}

We have established that all states of the signal mode in a photon-heralding experiment are nonclassical since their $P$-functions are negative at the origin. This is true even when the transmittances $\mu$ and $\eta$ are small, a situation where the environment interferes intensely with the system and where one may therefore have expected nonclassicality to be completely lost. We will investigate in this section how the nonclassicality of those states is quantitatively affected by the lowering of the transmittances away from their optimal value $\mu=\eta=1$ at different levels of the initial brightness $\xi$. For that purpose, we will use several measures of nonclassicality, discussed in the literature.

\subsection{Ordering sensitivity \label{sec:OS}}

Ordering sensitivity (OS) is a recently introduced measure of nonclassicality~\cite{Bievre19}, defined as the speed of change of the (second order) Renyi entropy of the $s$-ordered quasiprobability distribution with the ordering parameter $s$:
\begin{equation}\label{So}
\Os(\rho)=\frac\partial{\partial s}\left[\ln\left(\pi\int W^2(\alpha,s)d^2\alpha\right)\right]_{s=0}.
\end{equation}
The OS of all coherent states equals $1$ and all classical states have an OS less than $1$, which implies that
\begin{equation}\label{eq:wigner0criterium}
\Os(\rho)>1\Rightarrow \rho\ \text{is nonclassical}.
\end{equation}
The OS is therefore a nonclassicality witness. Note that it can be less than 1 for a nonclassical state. It is shown in~\cite{Bievre19} that a distance-based measure of nonclassicality can be constructed that is bounded between $\sqrt{\Os(\rho)}-1$ and $\sqrt{\Os(\rho)}$. Hence, for a sufficiently high $\Os(\rho)$, its square root is a good measure of nonclassicality.

An alternative expression that does not use the Wigner function is
\begin{equation}\label{So2}
\Os(\rho) = -\frac1{2\mathcal{P}(\rho)}\Tr\left(\left[Q,\rho\right]^2+\left[P,\rho\right]^2\right),
\end{equation}
where $Q=\frac1{\sqrt2}(a^\dagger +a)$ and $P=\frac{i}{\sqrt2}(a^\dagger-a)$ are two field quadratures, while $\mathcal{P}(\rho)=\Tr\rho^2$ is the purity of the state. For pure states OS coincides with the ``total noise'', the sum of variances of $Q$ and $P$ \cite{Hillery89}.

For a Fock-diagonal state $\rho=\sum_np_n|n\rangle\langle n|$, such as TDS, the OS is given by
\begin{equation}\label{Sodiag}
\Os(\rho) = \frac{\sum_n\left(p_n-p_{n+1}\right)^2(n+1)}{\sum_np_n^2}.
\end{equation}
Substituting the values of $p_k$ from Eq.~(\ref{rho-}) and summing up the geometric series, we obtain an analytic expression for the OS of TDS:
\begin{eqnarray}\label{Soqpd}
\Os(q, p, {d}) &=& \frac{\mathcal{C}^2}{\mathcal{P}(q, p, {d})}\left(\frac1{(1+q)^2}\right.\\\nonumber
&+& \left.\frac{{d}^2}{(1+qp)^2}-2d\frac{(1-q)(1-qp)}{(1-q^2p)^2}\right),
\end{eqnarray}
where the purity is
\begin{eqnarray}\label{Pqpd}
\mathcal{P}(q, p, {d}) &=& \mathcal{C}^2\left(\frac1{1-q^2}+\frac{{d}^2}{1-q^2p^2}-\frac{2d}{1-q^2p}\right).
\end{eqnarray}

Now we analyze the behaviour of the ordering sensitivity as function of the parameters of TDS. We first look at the situation where there are no losses in the idler and signal channels so that $\mu=1=\eta$ and consequently ${d}=1, p=0, q=\xi$. In this case the state of the signal mode is a truncated thermal state, see Sec.~\ref{sec:TTS}, and we find readily that
\begin{equation}\label{OsTTS}
\Os (q, 0, 1)=\frac{1-q}{1+q}(3+2q).
\end{equation}
One easily sees this quantity is maximal when $q=0$, where $\Os=3$, which is its value for the single photon state; it then decreases to $0$ as $q$ increases to $1$, which clearly illustrates the decrease of nonclassicality of the truncated thermal states as their temperature increases. Note that we know that all truncated thermal states are nonclassical since their $P$-function is negative at the origin $\alpha=0$. In fact, it is singular there, since it is the difference between a Gaussian and a delta function at the origin. Nevertheless, as this example shows, this negative singularity is not indicating a large ordering sensitivity of the states at high $q$. This can be further understood if one notices that the Wigner function of these states is also negative at the origin $\alpha=0$:
\begin{equation}\label{WigTTS}
W^{(-)}(0,0 \mid q, 1, 0)=-\frac{2}{\pi}\frac{1-q}{1+q}<0.
\end{equation}
However this negative value tends to zero as $q$ tends to $1$, also indicating loss of nonclassicality of the truncated thermal state with a growing temperature.

Let us now consider how the OS of TDS is affected if the transmittance $\mu$ of the signal mode is no longer maximal: $\mu<1$ but still $\eta=1$. Then $p=0$ and ${d}<1$ and Eq.~(\ref{Soqpd}) simplifies to
\begin{equation}\label{Osp0}
\Os (q, 0, {d})= \left(\frac{1-q}{1+q}\right)\frac{1+(1+q)^2({d}^2-2{d}(1-q))}{1+{d}({d}-2)(1-q^2)}.
\end{equation}

\begin{figure}[h!]
\includegraphics[width=\columnwidth]{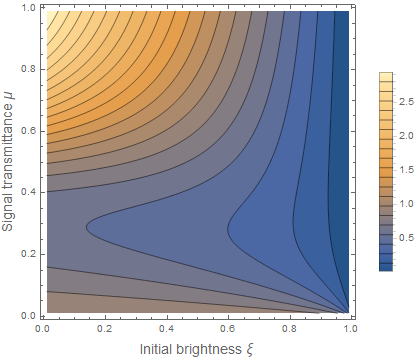}
\caption{Contour plot of the nonclassicality of TDS (measured by OS) for $\eta=1$ and ranges of $\xi$ and $\mu$ as shown. One notices that the maximum value for OS, $\OS=3$, is reached at $\mu=1$, $\xi\to0$, corresponding to the single-photon state. OS decreases with growing $\xi$ and/or diminishing $\mu$. However, at low values of $\mu<0.3$ OS grows again and has the second local maximum $\OS=1$ at $\mu\to0$, $\xi\to0$, corresponding to the vacuum state. \label{fig:OSmuqzero}}
\end{figure}

The contour plot of $\OS$ can be seen in Fig.~\ref{fig:OSmuqzero}. The maximal value of $\OS=3$ is reached for the single-photon state at $\mu=1$, $\xi\to0$. One clearly observes the rather fast loss of OS, and hence nonclassicality of the states with growing $\xi$ and/or diminishing $\mu$. For example, with $\xi$ as low as $0.2$ and $\mu$ as high as $0.85$ one sees that $\OS$ has decreased from its maximal value of $3$ to $1.99$.

It is interesting to note, that at low $\mu$ OS grows again and reaches the second local maximum $\OS=1$ at the vacuum state, which may be explained by growing purity as the state approaches the vacuum. This point is discussed in Sec.~\ref{sec:Comparison} below, where different nonclassicality measures are compared.

\begin{figure}[h!]
\includegraphics[width=\columnwidth]{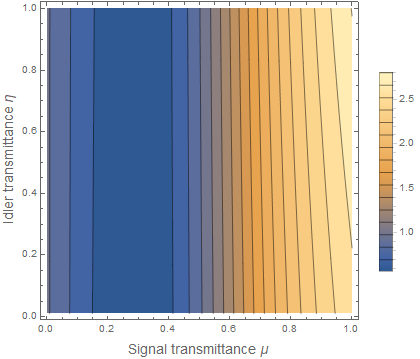}
\caption{Contour plot of the nonclassicality of TDS (measured by OS) for $\xi=0.05$ and ranges of $\eta$ and $\mu$ as shown. One notices that the OS is highly sensitive to losses in the signal channel and is almost insensitive to the losses in the idler one. \label{fig:OSmueta}}
\end{figure}

In Fig.~\ref{fig:OSmueta} we observe the dependence of the OS on both transmittances at low $\xi$. We see that the dependence on $\mu$ is very strong, while a weak dependence on $\eta$ exists only at high $\mu$. The origin of this dependence can be seen from the form taken by Eq.~(\ref{rho-}) in the limit $\xi\ll1$ and $\mu=1$, where we leave only terms linear in $\xi$:
\begin{equation}\label{rho-2}
\rho^{(-)} \approx \left[1-\xi(2-\eta)\right] |1\rangle\langle 1| + \xi(2-\eta)|2\rangle\langle 2|,
\end{equation}
showing clearly that the weight of the two-photon component decreases at high $\eta$ and the state becomes closer to the single-photon state, possessing the maximal non-classicality. On the other hand, keeping in the TDS, Eq.~(\ref{rho-}), only terms up to the first order in $\xi$ at any $\mu$, we find that the weight of the two-photon component is $\mu^2\xi(2-\eta)$, so that at low $\mu$ this weight becomes negligible. Thus, the quantum efficiency of the heralding detector, almost not affecting the nonclassicality at high signal loss, becomes important when the latter is small. A similar result was obtained recently in Ref.~\cite{Quesada19} for heralding experiments characterized by different figures of merit.

\begin{figure}
\includegraphics[width=0.8\columnwidth]{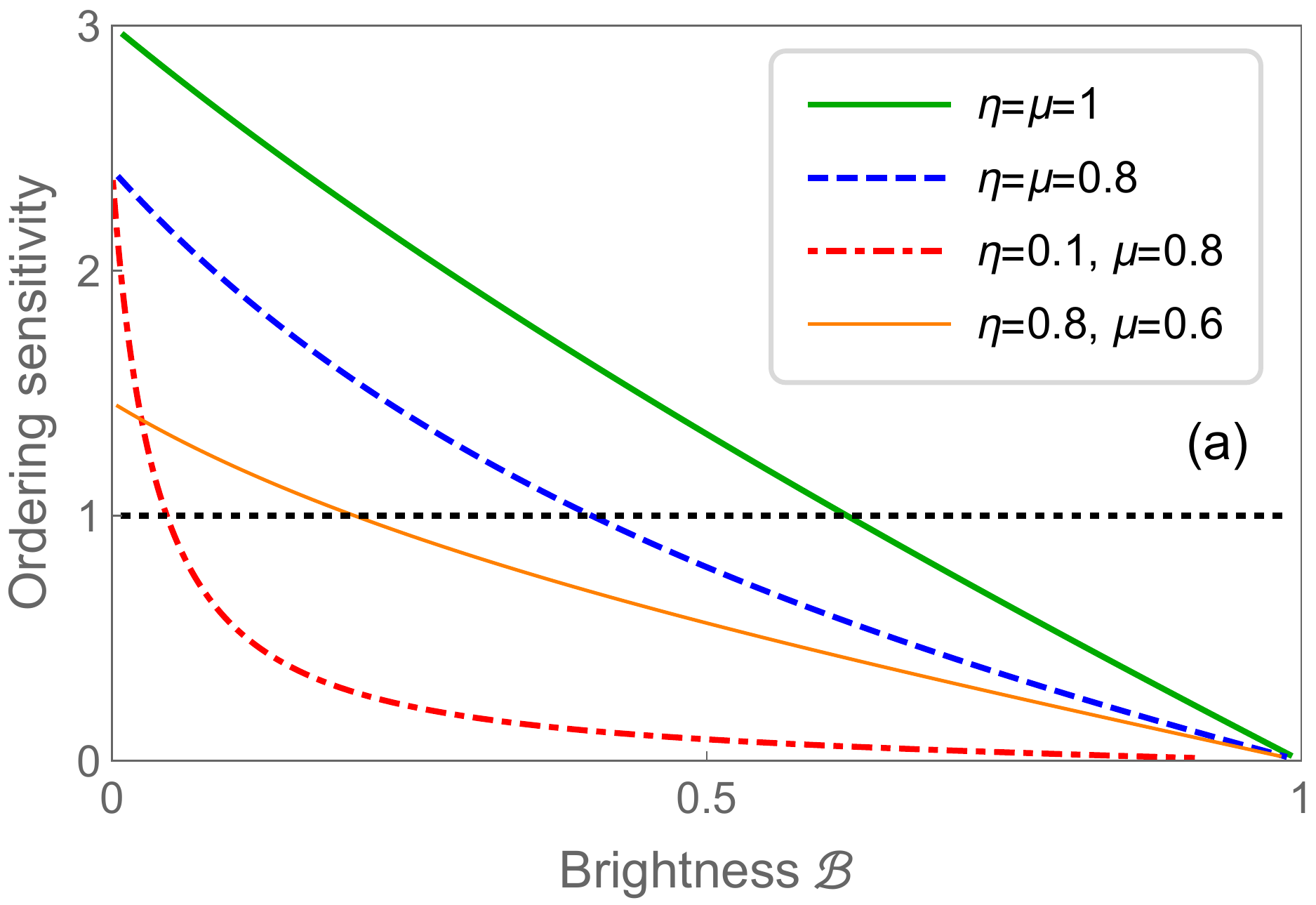}
\includegraphics[width=0.8\columnwidth]{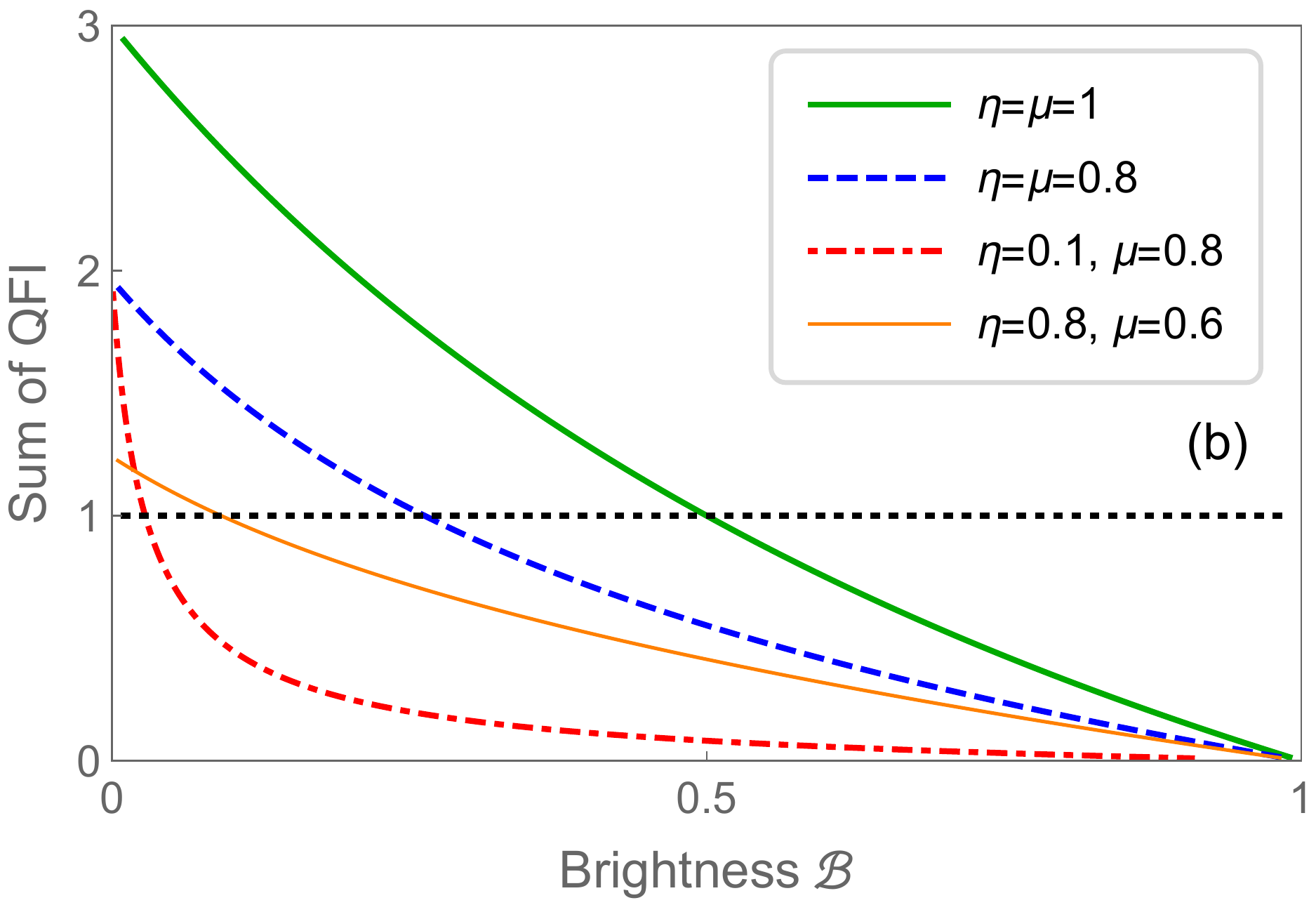}
\includegraphics[width=0.8\columnwidth]{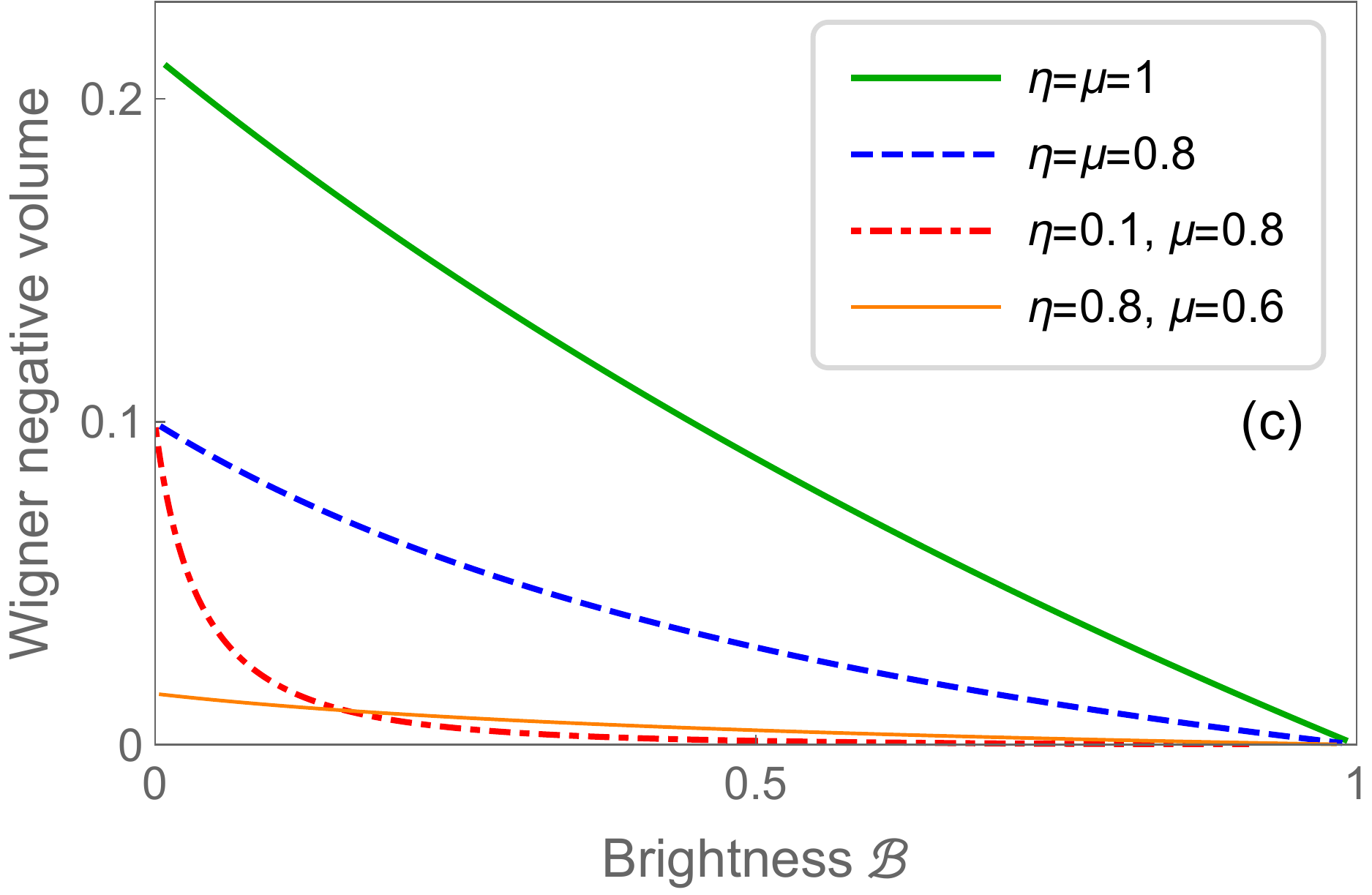}
\caption{Three witnesses of nonclassicality of TDS (serving also as measures of nonclassicality when they detect it) as functions of the brightness. Decreasing curves show a trade-off between the nonclassicality and the brightness, taking place at any level of losses. We see also that OS goes below 1 at higher values of brightness than the sum of QFI, meaning it is a better witness for this family of states. \label{fig:Nonclass}}
\end{figure}

We study also the dependence of the OS on the brightness $\mathcal{B}$, Eq.~(\ref{brightness}), for fixed values of the transmittances in both channels, Fig.~\ref{fig:Nonclass}(a). We see that the nonclassicality of the generated state always decreases with growing brightness. This fact represents a trade-off between the nonclassicality and the brightness similar to that between the single-photon purity and the brightness. At a realistic level of losses $\eta=\mu=0.8$ the nonclassicality becomes weak ($\OS<1$) at brightness higher than 0.4. If the losses of the signal mode raise to 50\%, the nonclassicality is weak at all levels of brightness. We conclude that highly nonclassical TDS are generated in the considered scheme at low initial brightness and low signal loss.

\subsection{Sum of quantum Fisher information}

Recently, a resource theory of nonclassicality has been proposed in which the measure of nonclassicality of a mixed state is the convex roof of the total noise of pure states, into which it can be decomposed~\cite{Yadin18, Kwon19}. Such convex roofs are however virtually impossible to compute, even on simple states. Useful lower bounds have been established using the quantum Fisher information (QFI) with respect to the quadrature $Q_\theta=Q\cos\theta + P\sin\theta$:

\begin{equation}\label{QFI}
\mathcal{F}(\rho,Q_\theta) = \frac12\sum_{a,b}\frac{\left(\lambda_a-\lambda_b\right)^2}{\lambda_a+\lambda_b} \left|\langle\psi_a|Q_\theta|\psi_b\rangle\right|^2,
\end{equation}
where $\rho=\sum_{a}\lambda_a|\psi_a\rangle\langle\psi_a|$ is a spectral decomposition and the sum is over all $a$, $b$ such that $\lambda_a+\lambda_b>0$. In our normalization for pure states $\mathcal{F}(\rho,Q_\theta)$ gives the variance of $Q_\theta$. A sum of QFI for two complementary quadratures
\begin{equation}\label{SQFI}
\mathcal{M}_\mathrm{QFI}(\rho) = \mathcal{F}(\rho,Q_\theta) + \mathcal{F}(\rho,Q_{\theta+\pi/2})
\end{equation}
is independent of the choice of $\theta$ and is an effective nonclassicality witness when $\mathcal{M}_\mathrm{QFI}(\rho)>1$. It is a lower bound for the convex roof of the total noise for pure states \cite{Yadin18}.

For a Fock-diagonal state $\rho=\sum_np_n|n\rangle\langle n|$, such as TDS, the QFI with respect to any quadrature is
\begin{equation}\label{QFIdiag}
\mathcal{F}(\rho,Q_\theta)  = \frac12\sum_n\frac{\left(p_n-p_{n+1}\right)^2}{p_n+p_{n+1}}(n+1),
\end{equation}
where the sum is over all $n$ such that $p_n+p_{n+1}>0$. For such states $\mathcal{M}_\mathrm{QFI}(\rho)$ is simply the double QFI with respect to any quadrature. Substituting the photon number distribution $p_k$ from Eq.~(\ref{rho-}) into Eq.~(\ref{QFIdiag}), we obtain after some algebra
\begin{eqnarray}\nonumber
\mathcal{M}_\mathrm{QFI}(\rho^{(-)}) &=& \mathcal{C}d\frac{(1-qp)^2}{1+qp}\left[(A_--A_+)^2\sum_{n=0}^\infty\frac{q^n(n+1)}{A_+-p^n}\right.\\\label{SQFI-} &+&\left.\frac{2A_--A_+}{(1-q)^2}-\frac1{(1-qp)^2}\right],
\end{eqnarray}
where $A_\pm=d^{-1}(1\pm q)/(1\pm qp)$. The infinite series in the last expression cannot be summed up analytically. However, the summation can be done up to some number $M-1$ and the remainder can be majorized by replacing $p^n$ in the denominator by its maximal value $p^M$. In this way we obtain
\begin{eqnarray}\nonumber
\mathcal{M}_\mathrm{QFI}^{(\mathrm{maj})}(\rho^{(-)}) &=& \mathcal{C}d\frac{(1-qp)^2}{1+qp}\left[(A_--A_+)^2\sum_{n=0}^{M-1}\frac{q^n(n+1)}{A_+-p^n}\right.\\\label{SQFI-m} &+&\left.\frac{(A_--A_+)^2}{A_+-p^M}\frac{M(1-q)+1}{(1-q)^2}q^M\right.\\\nonumber
&+&\left.\frac{2A_--A_+}{(1-q)^2}-\frac1{(1-qp)^2}\right] \ge \mathcal{M}_\mathrm{QFI}(\rho^{(-)}).
\end{eqnarray}
The dependence of $\mathcal{M}_\mathrm{QFI}^{(\mathrm{maj})}(\rho^{(-)})$ on the brightness of TDS is shown in Fig.~\ref{fig:Nonclass}(b). We see, that similar to the case of OS, considered in the previous section, the nonclassicality measured by the sum of QFI of two quadratures always decreases with growing brightness. The differences between the two measures are discussed in Sec.~\ref{sec:Comparison} below.

\subsection{Wigner negative volume}

The Wigner negative volume (WNV), defined as the absolute value of the integral of the Wigner function over the area where the latter is negative, is one more nonclassicality witness \cite{Kenfack04}, which we denote as $N_\mathrm{W}(\rho)$. Considering Eq.~(\ref{W-}) at $s=0$, we find that the Wigner function of a TDS is negative in a circle $|\alpha|<R$, where
\begin{equation}\label{R}
R^2 = \frac12\frac{\ln F}{\frac{1-qp}{1+qp}-\frac{1-q}{1+q}},
\end{equation}
with $F=d(1+q)/(1+qp)$, under condition that $F>1$. The absolute value of the integral of the Wigner function over this area is
\begin{equation}\label{Nw}
N_\mathrm{W}(\rho^{(-)}) = \left\{\begin{array}{ccc}
  A_0F^{-\frac{1-q^2p}{2q(1-p)}+\frac12} - 1, & \text{if} & F>1, \\
  0, & \text{if} & F\le1.\end{array}\right.
\end{equation}
where
\begin{equation}\label{A0}
A_0 = \frac{2q(1-p)}{(1-q)[1-qp-d(1-q)]}.
\end{equation}
In the limiting case $p=0$, $d=1$, $q\to0$ we find the WNV of the single-photon state:
\begin{equation}\label{Nw1}
N_\mathrm{W}(|1\rangle\langle1|) = \frac2{e^{1/2}}-1 \approx 0.213,
\end{equation}
which coincides with that found by a direct integration of the single-photon Wigner function \cite{Kenfack04}.

The dependence of $N_\mathrm{W}(\rho^{(-)})$ on the brightness of TDS is shown in Fig.~\ref{fig:Nonclass}(c). Again, as in two previous sections, the nonclassicality always decreases with growing brightness.

\subsection{Comparison of nonclassicality measures \label{sec:Comparison}}

Comparing Fig.~\ref{fig:Nonclass}(a) to Fig.~\ref{fig:Nonclass}(b), one notices that neither the OS nor the sum of QFI  detect the nonclassicality of the TDS perfectly, confirming that indeed, they are only nonclassicality witnesses and not nonclassicality measures. However the OS is more efficient as a witness for these states, since the region where it identifies their nonclassicality is larger. We recall that all states for $0<\mu\le1$ are nonclassical, and the nonclassicality is witnessed if $\OS>1$ or $\mathcal{M}_\mathrm{QFI}>1$. At $\eta=\mu=0.8$ the OS fails to witness the nonclassicality at brightness above 0.4, while the sum of QFI fails to do it for brightness above 0.26. Note, that the quantity shown in Fig.~\ref{fig:Nonclass}(b) is an upper bound, as indicated by Eq.~(\ref{SQFI-m}), and the true sum of QFI is even slightly lower. For other values of transmittances the situation is the same: the OS outperforms the sum of QFI. Both these witnesses, however, are outperformed by the WNV, which should be positive to witness the nonclassicality. We see, from Fig.~\ref{fig:Nonclass}(c) that the nonclassicality is always witnessed by the WNV for the considered examples. For other classes of states, like Gaussian states, the WNV is completely ineffective.

\begin{figure}[h!]
\includegraphics[width=\columnwidth]{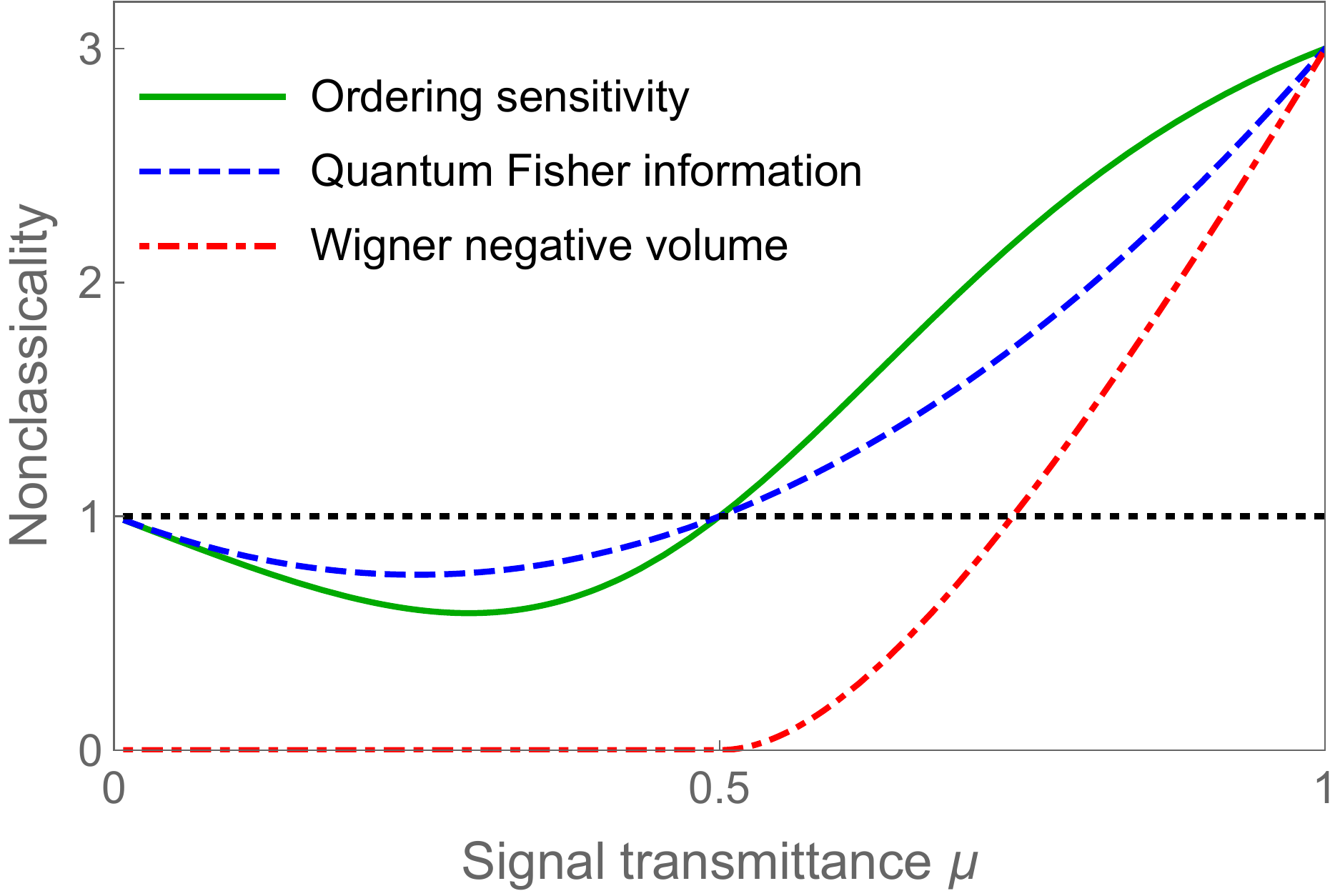}
\caption{Comparison of three nonclassicality witnesses for TDS with $\eta=1$ and $\xi=0.01$. The value for the Wigner negative volume is scaled to other two witnesses and is given by $3N_\mathrm{W}(\rho)/N_\mathrm{W}(|1\rangle\langle1|)$. In the region $0.5<\mu\le1$ all three witnesses detect the nonclassicality of the state, ascribing to it different measures. In the region $0<\mu\le0.5$ the OS and the sum of QFI may be growing with the increase of loss. However, in this region they ascribe no measure to the state. \label{fig:Comparison}}
\end{figure}

It is interesting also to compare different witnesses in the limiting case of $\eta=1$ and low $\xi$, where the TDS is close to a mixture of the vacuum and the single-photon state $(1-\mu)|0\rangle\langle0|+\mu|1\rangle\langle1|$, see Fig.~\ref{fig:Comparison}.

We see from Fig.~\ref{fig:Comparison}, that both the OS and the sum of QFI are growing with decreasing $\mu$, i.e. increasing loss, the fact which was mentioned above in Sec.~\ref{sec:OS}.  However, it happens in the region where both witnesses are less than 1 and do not detect nonclassicality. It is known from the resource theory of nonclassicality \cite{Yadin18,Kwon19}, that the sum of QFI is non-increasing (monotone) with loss in the region where it is higher than 1. Thus, its behaviour can be disregarded when it is below this limit. On the other hand, the square root of the OS, even if it is less than 1, provides an upper bound for a distance from the set of classical states \cite{Bievre19}. The growth of the OS with growing loss may mean that either the bound becomes looser with respect to the true value of the distance, or the set of classical states becomes sparser as one approaches the vacuum state by a trajectory determined by the considered family of states.

\section{Conclusions}

We have introduced a new three-parameter family of non-Gaussian single-mode optical states, the TDS. These states have non-singular $P$-functions, that can take negative values for some values of the parameters, that we identify. For those parameter values, the TDS are therefore nonclassical. We show furthermore that for these same values the TDS correspond to states of light, conditionally prepared by the technique of ``photon heralding''. In that context, the three parameters correspond to the losses in the signal and the idler modes and to the gain in the nonlinear crystal. In the absence of losses these states are known to be nonclassical. We have shown that this remains true for all values of the losses in both modes. We have shown that some well-known nonclassical states are members of this family at various limiting parameter values. A remarkable feature of this family of states is the possibility to obtain analytic expressions for various nonclassicality witnesses. We have calculated three such witnesses, the ordering sensitivity, the sum of quantum Fisher information for two quadratures and the Wigner negative volume and analyzed their dependencies on the three parameters. In particular, we have shown that the noclassicality of TDS is not very sensitive to the losses in the idler mode, but much more so to the losses of the signal one. We have established a general trade-off rule for the brightness and the nonclassicality of these states.

From the practical viewpoint, the nonclassical states of this family are non-Gaussian and represent thus a valuable resource for various protocols of quantum information processing. From the purely theoretical viewpoint, this family of states, whose quasiprobability distributions are differences of two Gaussians, are an excellent testbed for studying and comparing various measures of non-classicality, non-Gaussianity, etc.

A practical recommendation follows from the above analysis: in a photon-heralding experiment with some loss in the signal and the idler modes (always inevitable) it is much more efficient to reconstruct the $P$-function of the generated state, which is always regular and always negative, than the Wigner function, which may be positive for some combination of the experimental settings. Reconstruction of negative regions of the $P$-function will be a direct evidence of the state nonclassicality.

\section*{Acknowledgments}

DBH is grateful to Young-Sik Ra and Valery Shchesnovich for discussions. The authors thank both anonymous referees for their constructive comments that helped to improve the paper. This work was supported in part by the Labex CEMPI (ANR-11-LABX-0007-01) and by the Nord-Pas de Calais Regional Council and FEDER through the Contrat de Projets \'Etat-R\'egion (CPER), and in part by the European Union's Horizon 2020 research and innovation programme under grant agreement No 665148 (QCUMbER).

\bibliography{SinglePhoton2019b}

\begin{thebibliography}{62}%
\makeatletter
\providecommand \@ifxundefined [1]{%
 \@ifx{#1\undefined}
}%
\providecommand \@ifnum [1]{%
 \ifnum #1\expandafter \@firstoftwo
 \else \expandafter \@secondoftwo
 \fi
}%
\providecommand \@ifx [1]{%
 \ifx #1\expandafter \@firstoftwo
 \else \expandafter \@secondoftwo
 \fi
}%
\providecommand \natexlab [1]{#1}%
\providecommand \enquote  [1]{``#1''}%
\providecommand \bibnamefont  [1]{#1}%
\providecommand \bibfnamefont [1]{#1}%
\providecommand \citenamefont [1]{#1}%
\providecommand \href@noop [0]{\@secondoftwo}%
\providecommand \href [0]{\begingroup \@sanitize@url \@href}%
\providecommand \@href[1]{\@@startlink{#1}\@@href}%
\providecommand \@@href[1]{\endgroup#1\@@endlink}%
\providecommand \@sanitize@url [0]{\catcode `\\12\catcode `\$12\catcode
  `\&12\catcode `\#12\catcode `\^12\catcode `\_12\catcode `\%12\relax}%
\providecommand \@@startlink[1]{}%
\providecommand \@@endlink[0]{}%
\providecommand \url  [0]{\begingroup\@sanitize@url \@url }%
\providecommand \@url [1]{\endgroup\@href {#1}{\urlprefix }}%
\providecommand \urlprefix  [0]{URL }%
\providecommand \Eprint [0]{\href }%
\providecommand \doibase [0]{http://dx.doi.org/}%
\providecommand \selectlanguage [0]{\@gobble}%
\providecommand \bibinfo  [0]{\@secondoftwo}%
\providecommand \bibfield  [0]{\@secondoftwo}%
\providecommand \translation [1]{[#1]}%
\providecommand \BibitemOpen [0]{}%
\providecommand \bibitemStop [0]{}%
\providecommand \bibitemNoStop [0]{.\EOS\space}%
\providecommand \EOS [0]{\spacefactor3000\relax}%
\providecommand \BibitemShut  [1]{\csname bibitem#1\endcsname}%
\let\auto@bib@innerbib\@empty
\bibitem [{\citenamefont {Eisaman}\ \emph {et~al.}(2011)\citenamefont
  {Eisaman}, \citenamefont {Fan}, \citenamefont {Migdall},\ and\ \citenamefont
  {Polyakov}}]{Eisaman11}%
  \BibitemOpen
  \bibfield  {author} {\bibinfo {author} {\bibfnamefont {M.~D.}\ \bibnamefont
  {Eisaman}}, \bibinfo {author} {\bibfnamefont {J.}~\bibnamefont {Fan}},
  \bibinfo {author} {\bibfnamefont {A.}~\bibnamefont {Migdall}}, \ and\
  \bibinfo {author} {\bibfnamefont {S.~V.}\ \bibnamefont {Polyakov}},\ }\href
  {\doibase 10.1063/1.3610677} {\bibfield  {journal} {\bibinfo  {journal} {Rev.
  Sci. Instr.}\ }\textbf {\bibinfo {volume} {82}},\ \bibinfo {pages} {071101}
  (\bibinfo {year} {2011})}\BibitemShut {NoStop}%
\bibitem [{\citenamefont {Gisin}\ \emph {et~al.}(2002)\citenamefont {Gisin},
  \citenamefont {Ribordy}, \citenamefont {Tittel},\ and\ \citenamefont
  {Zbinden}}]{Gisin02}%
  \BibitemOpen
  \bibfield  {author} {\bibinfo {author} {\bibfnamefont {N.}~\bibnamefont
  {Gisin}}, \bibinfo {author} {\bibfnamefont {G.}~\bibnamefont {Ribordy}},
  \bibinfo {author} {\bibfnamefont {W.}~\bibnamefont {Tittel}}, \ and\ \bibinfo
  {author} {\bibfnamefont {H.}~\bibnamefont {Zbinden}},\ }\href {\doibase
  10.1103/RevModPhys.74.145} {\bibfield  {journal} {\bibinfo  {journal} {Rev.
  Mod. Phys.}\ }\textbf {\bibinfo {volume} {74}},\ \bibinfo {pages} {145}
  (\bibinfo {year} {2002})}\BibitemShut {NoStop}%
\bibitem [{\citenamefont {Sangouard}\ \emph {et~al.}(2011)\citenamefont
  {Sangouard}, \citenamefont {Simon}, \citenamefont {de~Riedmatten},\ and\
  \citenamefont {Gisin}}]{Sangouard11}%
  \BibitemOpen
  \bibfield  {author} {\bibinfo {author} {\bibfnamefont {N.}~\bibnamefont
  {Sangouard}}, \bibinfo {author} {\bibfnamefont {C.}~\bibnamefont {Simon}},
  \bibinfo {author} {\bibfnamefont {H.}~\bibnamefont {de~Riedmatten}}, \ and\
  \bibinfo {author} {\bibfnamefont {N.}~\bibnamefont {Gisin}},\ }\href
  {\doibase 10.1103/RevModPhys.83.33} {\bibfield  {journal} {\bibinfo
  {journal} {Rev. Mod. Phys.}\ }\textbf {\bibinfo {volume} {83}},\ \bibinfo
  {pages} {33} (\bibinfo {year} {2011})}\BibitemShut {NoStop}%
\bibitem [{\citenamefont {Knill}\ \emph {et~al.}(2001)\citenamefont {Knill},
  \citenamefont {Laflamme},\ and\ \citenamefont {Milburn}}]{Knill01}%
  \BibitemOpen
  \bibfield  {author} {\bibinfo {author} {\bibfnamefont {E.}~\bibnamefont
  {Knill}}, \bibinfo {author} {\bibfnamefont {R.}~\bibnamefont {Laflamme}}, \
  and\ \bibinfo {author} {\bibfnamefont {G.~J.}\ \bibnamefont {Milburn}},\
  }\href {\doibase 10.1038/35051009} {\bibfield  {journal} {\bibinfo  {journal}
  {Nature}\ }\textbf {\bibinfo {volume} {409}},\ \bibinfo {pages} {46}
  (\bibinfo {year} {2001})}\BibitemShut {NoStop}%
\bibitem [{\citenamefont {Kok}\ \emph {et~al.}(2007)\citenamefont {Kok},
  \citenamefont {Munro}, \citenamefont {Nemoto}, \citenamefont {Ralph},
  \citenamefont {Dowling},\ and\ \citenamefont {Milburn}}]{Kok07}%
  \BibitemOpen
  \bibfield  {author} {\bibinfo {author} {\bibfnamefont {P.}~\bibnamefont
  {Kok}}, \bibinfo {author} {\bibfnamefont {W.~J.}\ \bibnamefont {Munro}},
  \bibinfo {author} {\bibfnamefont {K.}~\bibnamefont {Nemoto}}, \bibinfo
  {author} {\bibfnamefont {T.~C.}\ \bibnamefont {Ralph}}, \bibinfo {author}
  {\bibfnamefont {J.~P.}\ \bibnamefont {Dowling}}, \ and\ \bibinfo {author}
  {\bibfnamefont {G.~J.}\ \bibnamefont {Milburn}},\ }\href {\doibase
  10.1103/RevModPhys.79.135} {\bibfield  {journal} {\bibinfo  {journal} {Rev.
  Mod. Phys.}\ }\textbf {\bibinfo {volume} {79}},\ \bibinfo {pages} {135}
  (\bibinfo {year} {2007})}\BibitemShut {NoStop}%
\bibitem [{\citenamefont {Tillmann}\ \emph {et~al.}(2013)\citenamefont
  {Tillmann}, \citenamefont {Daki{\'c}}, \citenamefont {Heilmann},
  \citenamefont {Nolte}, \citenamefont {Szameit},\ and\ \citenamefont
  {Walther}}]{Tillmann13}%
  \BibitemOpen
  \bibfield  {author} {\bibinfo {author} {\bibfnamefont {M.}~\bibnamefont
  {Tillmann}}, \bibinfo {author} {\bibfnamefont {B.}~\bibnamefont {Daki{\'c}}},
  \bibinfo {author} {\bibfnamefont {R.}~\bibnamefont {Heilmann}}, \bibinfo
  {author} {\bibfnamefont {S.}~\bibnamefont {Nolte}}, \bibinfo {author}
  {\bibfnamefont {A.}~\bibnamefont {Szameit}}, \ and\ \bibinfo {author}
  {\bibfnamefont {P.}~\bibnamefont {Walther}},\ }\href@noop {} {\bibfield
  {journal} {\bibinfo  {journal} {Nat. Photonics}\ }\textbf {\bibinfo {volume}
  {7}},\ \bibinfo {pages} {540} (\bibinfo {year} {2013})}\BibitemShut {NoStop}%
\bibitem [{\citenamefont {Shchesnovich}(2014)}]{Shchesnovich14}%
  \BibitemOpen
  \bibfield  {author} {\bibinfo {author} {\bibfnamefont {V.}~\bibnamefont
  {Shchesnovich}},\ }\href@noop {} {\bibfield  {journal} {\bibinfo  {journal}
  {Phys. Rev. A}\ }\textbf {\bibinfo {volume} {89}},\ \bibinfo {pages} {022333}
  (\bibinfo {year} {2014})}\BibitemShut {NoStop}%
\bibitem [{\citenamefont {Quesada}\ \emph {et~al.}(2018)\citenamefont
  {Quesada}, \citenamefont {Arrazola},\ and\ \citenamefont
  {Killoran}}]{Quesada18}%
  \BibitemOpen
  \bibfield  {author} {\bibinfo {author} {\bibfnamefont {N.}~\bibnamefont
  {Quesada}}, \bibinfo {author} {\bibfnamefont {J.~M.}\ \bibnamefont
  {Arrazola}}, \ and\ \bibinfo {author} {\bibfnamefont {N.}~\bibnamefont
  {Killoran}},\ }\href {\doibase 10.1103/PhysRevA.98.062322} {\bibfield
  {journal} {\bibinfo  {journal} {Phys. Rev. A}\ }\textbf {\bibinfo {volume}
  {98}},\ \bibinfo {pages} {062322} (\bibinfo {year} {2018})}\BibitemShut
  {NoStop}%
\bibitem [{\citenamefont {Klyshko}(1988)}]{KlyshkoBook}%
  \BibitemOpen
  \bibfield  {author} {\bibinfo {author} {\bibfnamefont {D.~N.}\ \bibnamefont
  {Klyshko}},\ }\href@noop {} {\emph {\bibinfo {title} {Photons and Nonlinear
  Optics}}}\ (\bibinfo  {publisher} {Gordon and Breach},\ \bibinfo {address}
  {New York},\ \bibinfo {year} {1988})\BibitemShut {NoStop}%
\bibitem [{\citenamefont {Brida}\ \emph {et~al.}(2006)\citenamefont {Brida},
  \citenamefont {Genovese},\ and\ \citenamefont {Gramegna}}]{Brida06}%
  \BibitemOpen
  \bibfield  {author} {\bibinfo {author} {\bibfnamefont {G.}~\bibnamefont
  {Brida}}, \bibinfo {author} {\bibfnamefont {M.}~\bibnamefont {Genovese}}, \
  and\ \bibinfo {author} {\bibfnamefont {M.}~\bibnamefont {Gramegna}},\ }\href
  {http://stacks.iop.org/1612-202X/3/i=3/a=001} {\bibfield  {journal} {\bibinfo
   {journal} {Laser Phys. Lett.}\ }\textbf {\bibinfo {volume} {3}},\ \bibinfo
  {pages} {115} (\bibinfo {year} {2006})}\BibitemShut {NoStop}%
\bibitem [{\citenamefont {Rodiek}\ \emph {et~al.}(2017)\citenamefont {Rodiek},
  \citenamefont {Lopez}, \citenamefont {Hofer}, \citenamefont {Porrovecchio},
  \citenamefont {Smid}, \citenamefont {Chu}, \citenamefont {Gotzinger},
  \citenamefont {Sandoghdar}, \citenamefont {Lindner}, \citenamefont {Becher},\
  and\ \citenamefont {Kuck}}]{Rodiek17}%
  \BibitemOpen
  \bibfield  {author} {\bibinfo {author} {\bibfnamefont {B.}~\bibnamefont
  {Rodiek}}, \bibinfo {author} {\bibfnamefont {M.}~\bibnamefont {Lopez}},
  \bibinfo {author} {\bibfnamefont {H.}~\bibnamefont {Hofer}}, \bibinfo
  {author} {\bibfnamefont {G.}~\bibnamefont {Porrovecchio}}, \bibinfo {author}
  {\bibfnamefont {M.}~\bibnamefont {Smid}}, \bibinfo {author} {\bibfnamefont
  {X.-L.}\ \bibnamefont {Chu}}, \bibinfo {author} {\bibfnamefont
  {S.}~\bibnamefont {Gotzinger}}, \bibinfo {author} {\bibfnamefont
  {V.}~\bibnamefont {Sandoghdar}}, \bibinfo {author} {\bibfnamefont
  {S.}~\bibnamefont {Lindner}}, \bibinfo {author} {\bibfnamefont
  {C.}~\bibnamefont {Becher}}, \ and\ \bibinfo {author} {\bibfnamefont
  {S.}~\bibnamefont {Kuck}},\ }\href {\doibase 10.1364/OPTICA.4.000071}
  {\bibfield  {journal} {\bibinfo  {journal} {Optica}\ }\textbf {\bibinfo
  {volume} {4}},\ \bibinfo {pages} {71} (\bibinfo {year} {2017})}\BibitemShut
  {NoStop}%
\bibitem [{\citenamefont {Somaschi}\ \emph {et~al.}(2016)\citenamefont
  {Somaschi}, \citenamefont {Giesz}, \citenamefont {De~Santis}, \citenamefont
  {Loredo}, \citenamefont {Almeida}, \citenamefont {Hornecker}, \citenamefont
  {Portalupi}, \citenamefont {Grange}, \citenamefont {Ant{\'o}n}, \citenamefont
  {Demory} \emph {et~al.}}]{Somaschi16}%
  \BibitemOpen
  \bibfield  {author} {\bibinfo {author} {\bibfnamefont {N.}~\bibnamefont
  {Somaschi}}, \bibinfo {author} {\bibfnamefont {V.}~\bibnamefont {Giesz}},
  \bibinfo {author} {\bibfnamefont {L.}~\bibnamefont {De~Santis}}, \bibinfo
  {author} {\bibfnamefont {J.}~\bibnamefont {Loredo}}, \bibinfo {author}
  {\bibfnamefont {M.~P.}\ \bibnamefont {Almeida}}, \bibinfo {author}
  {\bibfnamefont {G.}~\bibnamefont {Hornecker}}, \bibinfo {author}
  {\bibfnamefont {S.~L.}\ \bibnamefont {Portalupi}}, \bibinfo {author}
  {\bibfnamefont {T.}~\bibnamefont {Grange}}, \bibinfo {author} {\bibfnamefont
  {C.}~\bibnamefont {Ant{\'o}n}}, \bibinfo {author} {\bibfnamefont
  {J.}~\bibnamefont {Demory}},  \emph {et~al.},\ }\href@noop {} {\bibfield
  {journal} {\bibinfo  {journal} {Nat. Photonics}\ }\textbf {\bibinfo {volume}
  {10}},\ \bibinfo {pages} {340} (\bibinfo {year} {2016})}\BibitemShut
  {NoStop}%
\bibitem [{\citenamefont {Ding}\ \emph {et~al.}(2016)\citenamefont {Ding},
  \citenamefont {He}, \citenamefont {Duan}, \citenamefont {Gregersen},
  \citenamefont {Chen}, \citenamefont {Unsleber}, \citenamefont {Maier},
  \citenamefont {Schneider}, \citenamefont {Kamp}, \citenamefont {H\"ofling},
  \citenamefont {Lu},\ and\ \citenamefont {Pan}}]{Ding16}%
  \BibitemOpen
  \bibfield  {author} {\bibinfo {author} {\bibfnamefont {X.}~\bibnamefont
  {Ding}}, \bibinfo {author} {\bibfnamefont {Y.}~\bibnamefont {He}}, \bibinfo
  {author} {\bibfnamefont {Z.-C.}\ \bibnamefont {Duan}}, \bibinfo {author}
  {\bibfnamefont {N.}~\bibnamefont {Gregersen}}, \bibinfo {author}
  {\bibfnamefont {M.-C.}\ \bibnamefont {Chen}}, \bibinfo {author}
  {\bibfnamefont {S.}~\bibnamefont {Unsleber}}, \bibinfo {author}
  {\bibfnamefont {S.}~\bibnamefont {Maier}}, \bibinfo {author} {\bibfnamefont
  {C.}~\bibnamefont {Schneider}}, \bibinfo {author} {\bibfnamefont
  {M.}~\bibnamefont {Kamp}}, \bibinfo {author} {\bibfnamefont {S.}~\bibnamefont
  {H\"ofling}}, \bibinfo {author} {\bibfnamefont {C.-Y.}\ \bibnamefont {Lu}}, \
  and\ \bibinfo {author} {\bibfnamefont {J.-W.}\ \bibnamefont {Pan}},\ }\href
  {\doibase 10.1103/PhysRevLett.116.020401} {\bibfield  {journal} {\bibinfo
  {journal} {Phys. Rev. Lett.}\ }\textbf {\bibinfo {volume} {116}},\ \bibinfo
  {pages} {020401} (\bibinfo {year} {2016})}\BibitemShut {NoStop}%
\bibitem [{\citenamefont {Fedotov}\ \emph {et~al.}(2012)\citenamefont
  {Fedotov}, \citenamefont {Safronov}, \citenamefont {Shandarov}, \citenamefont
  {Lanin}, \citenamefont {Fedotov}, \citenamefont {Kilin}, \citenamefont
  {Sakoda}, \citenamefont {Scully},\ and\ \citenamefont
  {Zheltikov}}]{Fedotov12}%
  \BibitemOpen
  \bibfield  {author} {\bibinfo {author} {\bibfnamefont {I.~V.}\ \bibnamefont
  {Fedotov}}, \bibinfo {author} {\bibfnamefont {N.~A.}\ \bibnamefont
  {Safronov}}, \bibinfo {author} {\bibfnamefont {Y.~A.}\ \bibnamefont
  {Shandarov}}, \bibinfo {author} {\bibfnamefont {A.~A.}\ \bibnamefont
  {Lanin}}, \bibinfo {author} {\bibfnamefont {A.~B.}\ \bibnamefont {Fedotov}},
  \bibinfo {author} {\bibfnamefont {S.~Y.}\ \bibnamefont {Kilin}}, \bibinfo
  {author} {\bibfnamefont {K.}~\bibnamefont {Sakoda}}, \bibinfo {author}
  {\bibfnamefont {M.~O.}\ \bibnamefont {Scully}}, \ and\ \bibinfo {author}
  {\bibfnamefont {A.~M.}\ \bibnamefont {Zheltikov}},\ }\href {\doibase
  10.1063/1.4731762} {\bibfield  {journal} {\bibinfo  {journal} {Appl. Phys.
  Lett.}\ }\textbf {\bibinfo {volume} {101}},\ \bibinfo {pages} {031106}
  (\bibinfo {year} {2012})}\BibitemShut {NoStop}%
\bibitem [{\citenamefont {Sipahigil}\ \emph {et~al.}(2014)\citenamefont
  {Sipahigil}, \citenamefont {Jahnke}, \citenamefont {Rogers}, \citenamefont
  {Teraji}, \citenamefont {Isoya}, \citenamefont {Zibrov}, \citenamefont
  {Jelezko},\ and\ \citenamefont {Lukin}}]{Sipahigil14}%
  \BibitemOpen
  \bibfield  {author} {\bibinfo {author} {\bibfnamefont {A.}~\bibnamefont
  {Sipahigil}}, \bibinfo {author} {\bibfnamefont {K.~D.}\ \bibnamefont
  {Jahnke}}, \bibinfo {author} {\bibfnamefont {L.~J.}\ \bibnamefont {Rogers}},
  \bibinfo {author} {\bibfnamefont {T.}~\bibnamefont {Teraji}}, \bibinfo
  {author} {\bibfnamefont {J.}~\bibnamefont {Isoya}}, \bibinfo {author}
  {\bibfnamefont {A.~S.}\ \bibnamefont {Zibrov}}, \bibinfo {author}
  {\bibfnamefont {F.}~\bibnamefont {Jelezko}}, \ and\ \bibinfo {author}
  {\bibfnamefont {M.~D.}\ \bibnamefont {Lukin}},\ }\href {\doibase
  10.1103/PhysRevLett.113.113602} {\bibfield  {journal} {\bibinfo  {journal}
  {Phys. Rev. Lett.}\ }\textbf {\bibinfo {volume} {113}},\ \bibinfo {pages}
  {113602} (\bibinfo {year} {2014})}\BibitemShut {NoStop}%
\bibitem [{\citenamefont {Chu}\ \emph {et~al.}(2017)\citenamefont {Chu},
  \citenamefont {G{\"o}tzinger},\ and\ \citenamefont {Sandoghdar}}]{Chu17}%
  \BibitemOpen
  \bibfield  {author} {\bibinfo {author} {\bibfnamefont {X.-L.}\ \bibnamefont
  {Chu}}, \bibinfo {author} {\bibfnamefont {S.}~\bibnamefont {G{\"o}tzinger}},
  \ and\ \bibinfo {author} {\bibfnamefont {V.}~\bibnamefont {Sandoghdar}},\
  }\href@noop {} {\bibfield  {journal} {\bibinfo  {journal} {Nat. Photonics}\
  }\textbf {\bibinfo {volume} {11}},\ \bibinfo {pages} {58} (\bibinfo {year}
  {2017})}\BibitemShut {NoStop}%
\bibitem [{\citenamefont {Rezai}\ \emph {et~al.}(2018)\citenamefont {Rezai},
  \citenamefont {Wrachtrup},\ and\ \citenamefont {Gerhardt}}]{Rezai18}%
  \BibitemOpen
  \bibfield  {author} {\bibinfo {author} {\bibfnamefont {M.}~\bibnamefont
  {Rezai}}, \bibinfo {author} {\bibfnamefont {J.}~\bibnamefont {Wrachtrup}}, \
  and\ \bibinfo {author} {\bibfnamefont {I.}~\bibnamefont {Gerhardt}},\ }\href
  {\doibase 10.1103/PhysRevX.8.031026} {\bibfield  {journal} {\bibinfo
  {journal} {Phys. Rev. X}\ }\textbf {\bibinfo {volume} {8}},\ \bibinfo {pages}
  {031026} (\bibinfo {year} {2018})}\BibitemShut {NoStop}%
\bibitem [{\citenamefont {Kilin}\ and\ \citenamefont
  {Horoshko}(1995)}]{Kilin95}%
  \BibitemOpen
  \bibfield  {author} {\bibinfo {author} {\bibfnamefont {S.~Y.}\ \bibnamefont
  {Kilin}}\ and\ \bibinfo {author} {\bibfnamefont {D.~B.}\ \bibnamefont
  {Horoshko}},\ }\href {\doibase 10.1103/PhysRevLett.74.5206} {\bibfield
  {journal} {\bibinfo  {journal} {Phys. Rev. Lett.}\ }\textbf {\bibinfo
  {volume} {74}},\ \bibinfo {pages} {5206} (\bibinfo {year}
  {1995})}\BibitemShut {NoStop}%
\bibitem [{\citenamefont {Mogilevtsev}\ and\ \citenamefont
  {Shchesnovich}(2010)}]{Mogilevtsev10}%
  \BibitemOpen
  \bibfield  {author} {\bibinfo {author} {\bibfnamefont {D.}~\bibnamefont
  {Mogilevtsev}}\ and\ \bibinfo {author} {\bibfnamefont {V.~S.}\ \bibnamefont
  {Shchesnovich}},\ }\href {\doibase 10.1364/OL.35.003375} {\bibfield
  {journal} {\bibinfo  {journal} {Opt. Lett.}\ }\textbf {\bibinfo {volume}
  {35}},\ \bibinfo {pages} {3375} (\bibinfo {year} {2010})}\BibitemShut
  {NoStop}%
\bibitem [{\citenamefont {{Zel'dovich}}\ and\ \citenamefont
  {{Klyshko}}(1969)}]{Zeldovich69}%
  \BibitemOpen
  \bibfield  {author} {\bibinfo {author} {\bibfnamefont {B.~Y.}\ \bibnamefont
  {{Zel'dovich}}}\ and\ \bibinfo {author} {\bibfnamefont {D.~N.}\ \bibnamefont
  {{Klyshko}}},\ }\href@noop {} {\bibfield  {journal} {\bibinfo  {journal}
  {JETP Lett.}\ }\textbf {\bibinfo {volume} {9}},\ \bibinfo {pages} {40}
  (\bibinfo {year} {1969})}\BibitemShut {NoStop}%
\bibitem [{\citenamefont {Hong}\ and\ \citenamefont {Mandel}(1986)}]{Hong86}%
  \BibitemOpen
  \bibfield  {author} {\bibinfo {author} {\bibfnamefont {C.~K.}\ \bibnamefont
  {Hong}}\ and\ \bibinfo {author} {\bibfnamefont {L.}~\bibnamefont {Mandel}},\
  }\href {\doibase 10.1103/PhysRevLett.56.58} {\bibfield  {journal} {\bibinfo
  {journal} {Phys. Rev. Lett.}\ }\textbf {\bibinfo {volume} {56}},\ \bibinfo
  {pages} {58} (\bibinfo {year} {1986})}\BibitemShut {NoStop}%
\bibitem [{\citenamefont {Grangier}\ \emph {et~al.}(1986)\citenamefont
  {Grangier}, \citenamefont {Roger},\ and\ \citenamefont
  {Aspect}}]{Grangier86}%
  \BibitemOpen
  \bibfield  {author} {\bibinfo {author} {\bibfnamefont {P.}~\bibnamefont
  {Grangier}}, \bibinfo {author} {\bibfnamefont {G.}~\bibnamefont {Roger}}, \
  and\ \bibinfo {author} {\bibfnamefont {A.}~\bibnamefont {Aspect}},\ }\href
  {http://stacks.iop.org/0295-5075/1/i=4/a=004} {\bibfield  {journal} {\bibinfo
   {journal} {Europhys. Lett.}\ }\textbf {\bibinfo {volume} {1}},\ \bibinfo
  {pages} {173} (\bibinfo {year} {1986})}\BibitemShut {NoStop}%
\bibitem [{\citenamefont {Lvovsky}\ \emph {et~al.}(2001)\citenamefont
  {Lvovsky}, \citenamefont {Hansen}, \citenamefont {Aichele}, \citenamefont
  {Benson}, \citenamefont {Mlynek},\ and\ \citenamefont
  {Schiller}}]{Lvovsky01}%
  \BibitemOpen
  \bibfield  {author} {\bibinfo {author} {\bibfnamefont {A.~I.}\ \bibnamefont
  {Lvovsky}}, \bibinfo {author} {\bibfnamefont {H.}~\bibnamefont {Hansen}},
  \bibinfo {author} {\bibfnamefont {T.}~\bibnamefont {Aichele}}, \bibinfo
  {author} {\bibfnamefont {O.}~\bibnamefont {Benson}}, \bibinfo {author}
  {\bibfnamefont {J.}~\bibnamefont {Mlynek}}, \ and\ \bibinfo {author}
  {\bibfnamefont {S.}~\bibnamefont {Schiller}},\ }\href {\doibase
  10.1103/PhysRevLett.87.050402} {\bibfield  {journal} {\bibinfo  {journal}
  {Phys. Rev. Lett.}\ }\textbf {\bibinfo {volume} {87}},\ \bibinfo {pages}
  {050402} (\bibinfo {year} {2001})}\BibitemShut {NoStop}%
\bibitem [{\citenamefont {U'Ren}\ \emph {et~al.}(2004)\citenamefont {U'Ren},
  \citenamefont {Silberhorn}, \citenamefont {Banaszek},\ and\ \citenamefont
  {Walmsley}}]{URen04}%
  \BibitemOpen
  \bibfield  {author} {\bibinfo {author} {\bibfnamefont {A.~B.}\ \bibnamefont
  {U'Ren}}, \bibinfo {author} {\bibfnamefont {C.}~\bibnamefont {Silberhorn}},
  \bibinfo {author} {\bibfnamefont {K.}~\bibnamefont {Banaszek}}, \ and\
  \bibinfo {author} {\bibfnamefont {I.~A.}\ \bibnamefont {Walmsley}},\ }\href
  {\doibase 10.1103/PhysRevLett.93.093601} {\bibfield  {journal} {\bibinfo
  {journal} {Phys. Rev. Lett.}\ }\textbf {\bibinfo {volume} {93}},\ \bibinfo
  {pages} {093601} (\bibinfo {year} {2004})}\BibitemShut {NoStop}%
\bibitem [{\citenamefont {Fasel}\ \emph {et~al.}(2004)\citenamefont {Fasel},
  \citenamefont {Alibart}, \citenamefont {Tanzilli}, \citenamefont {Baldi},
  \citenamefont {Beveratos}, \citenamefont {Gisin},\ and\ \citenamefont
  {Zbinden}}]{Fasel04}%
  \BibitemOpen
  \bibfield  {author} {\bibinfo {author} {\bibfnamefont {S.}~\bibnamefont
  {Fasel}}, \bibinfo {author} {\bibfnamefont {O.}~\bibnamefont {Alibart}},
  \bibinfo {author} {\bibfnamefont {S.}~\bibnamefont {Tanzilli}}, \bibinfo
  {author} {\bibfnamefont {P.}~\bibnamefont {Baldi}}, \bibinfo {author}
  {\bibfnamefont {A.}~\bibnamefont {Beveratos}}, \bibinfo {author}
  {\bibfnamefont {N.}~\bibnamefont {Gisin}}, \ and\ \bibinfo {author}
  {\bibfnamefont {H.}~\bibnamefont {Zbinden}},\ }\href {\doibase
  10.1088/1367-2630/6/1/163} {\bibfield  {journal} {\bibinfo  {journal} {New J.
  Phys.}\ }\textbf {\bibinfo {volume} {6}},\ \bibinfo {pages} {163} (\bibinfo
  {year} {2004})}\BibitemShut {NoStop}%
\bibitem [{\citenamefont {Neergaard-Nielsen}\ \emph {et~al.}(2007)\citenamefont
  {Neergaard-Nielsen}, \citenamefont {Nielsen}, \citenamefont {Takahashi},
  \citenamefont {Vistnes},\ and\ \citenamefont {Polzik}}]{Neergaard-Nielsen07}%
  \BibitemOpen
  \bibfield  {author} {\bibinfo {author} {\bibfnamefont {J.~S.}\ \bibnamefont
  {Neergaard-Nielsen}}, \bibinfo {author} {\bibfnamefont {B.~M.}\ \bibnamefont
  {Nielsen}}, \bibinfo {author} {\bibfnamefont {H.}~\bibnamefont {Takahashi}},
  \bibinfo {author} {\bibfnamefont {A.~I.}\ \bibnamefont {Vistnes}}, \ and\
  \bibinfo {author} {\bibfnamefont {E.~S.}\ \bibnamefont {Polzik}},\ }\href
  {\doibase 10.1364/OE.15.007940} {\bibfield  {journal} {\bibinfo  {journal}
  {Opt. Express}\ }\textbf {\bibinfo {volume} {15}},\ \bibinfo {pages} {7940}
  (\bibinfo {year} {2007})}\BibitemShut {NoStop}%
\bibitem [{\citenamefont {Brida}\ \emph {et~al.}(2011)\citenamefont {Brida},
  \citenamefont {Degiovanni}, \citenamefont {Genovese}, \citenamefont
  {Migdall}, \citenamefont {Piacentini}, \citenamefont {Polyakov},\ and\
  \citenamefont {Berchera}}]{Brida11}%
  \BibitemOpen
  \bibfield  {author} {\bibinfo {author} {\bibfnamefont {G.}~\bibnamefont
  {Brida}}, \bibinfo {author} {\bibfnamefont {I.~P.}\ \bibnamefont
  {Degiovanni}}, \bibinfo {author} {\bibfnamefont {M.}~\bibnamefont
  {Genovese}}, \bibinfo {author} {\bibfnamefont {A.}~\bibnamefont {Migdall}},
  \bibinfo {author} {\bibfnamefont {F.}~\bibnamefont {Piacentini}}, \bibinfo
  {author} {\bibfnamefont {S.~V.}\ \bibnamefont {Polyakov}}, \ and\ \bibinfo
  {author} {\bibfnamefont {I.~R.}\ \bibnamefont {Berchera}},\ }\href {\doibase
  10.1364/OE.19.001484} {\bibfield  {journal} {\bibinfo  {journal} {Opt.
  Express}\ }\textbf {\bibinfo {volume} {19}},\ \bibinfo {pages} {1484}
  (\bibinfo {year} {2011})}\BibitemShut {NoStop}%
\bibitem [{\citenamefont {F{\"o}rtsch}\ \emph {et~al.}(2013)\citenamefont
  {F{\"o}rtsch}, \citenamefont {F{\"u}rst}, \citenamefont {Wittmann},
  \citenamefont {Strekalov}, \citenamefont {Aiello}, \citenamefont {Chekhova},
  \citenamefont {Silberhorn}, \citenamefont {Leuchs},\ and\ \citenamefont
  {Marquardt}}]{Fortsch13}%
  \BibitemOpen
  \bibfield  {author} {\bibinfo {author} {\bibfnamefont {M.}~\bibnamefont
  {F{\"o}rtsch}}, \bibinfo {author} {\bibfnamefont {J.~U.}\ \bibnamefont
  {F{\"u}rst}}, \bibinfo {author} {\bibfnamefont {C.}~\bibnamefont {Wittmann}},
  \bibinfo {author} {\bibfnamefont {D.}~\bibnamefont {Strekalov}}, \bibinfo
  {author} {\bibfnamefont {A.}~\bibnamefont {Aiello}}, \bibinfo {author}
  {\bibfnamefont {M.~V.}\ \bibnamefont {Chekhova}}, \bibinfo {author}
  {\bibfnamefont {C.}~\bibnamefont {Silberhorn}}, \bibinfo {author}
  {\bibfnamefont {G.}~\bibnamefont {Leuchs}}, \ and\ \bibinfo {author}
  {\bibfnamefont {C.}~\bibnamefont {Marquardt}},\ }\href@noop {} {\bibfield
  {journal} {\bibinfo  {journal} {Nat. Commun.}\ }\textbf {\bibinfo {volume}
  {4}},\ \bibinfo {pages} {1818} (\bibinfo {year} {2013})}\BibitemShut
  {NoStop}%
\bibitem [{\citenamefont {Kaneda}\ \emph {et~al.}(2015)\citenamefont {Kaneda},
  \citenamefont {Christensen}, \citenamefont {Wong}, \citenamefont {Park},
  \citenamefont {McCusker},\ and\ \citenamefont {Kwiat}}]{Kaneda15}%
  \BibitemOpen
  \bibfield  {author} {\bibinfo {author} {\bibfnamefont {F.}~\bibnamefont
  {Kaneda}}, \bibinfo {author} {\bibfnamefont {B.~G.}\ \bibnamefont
  {Christensen}}, \bibinfo {author} {\bibfnamefont {J.~J.}\ \bibnamefont
  {Wong}}, \bibinfo {author} {\bibfnamefont {H.~S.}\ \bibnamefont {Park}},
  \bibinfo {author} {\bibfnamefont {K.~T.}\ \bibnamefont {McCusker}}, \ and\
  \bibinfo {author} {\bibfnamefont {P.~G.}\ \bibnamefont {Kwiat}},\ }\href
  {\doibase 10.1364/OPTICA.2.001010} {\bibfield  {journal} {\bibinfo  {journal}
  {Optica}\ }\textbf {\bibinfo {volume} {2}},\ \bibinfo {pages} {1010}
  (\bibinfo {year} {2015})}\BibitemShut {NoStop}%
\bibitem [{\citenamefont {Joshi}\ \emph {et~al.}(2018)\citenamefont {Joshi},
  \citenamefont {Farsi}, \citenamefont {Clemmen}, \citenamefont {Ramelow},\
  and\ \citenamefont {Gaeta}}]{Joshi18}%
  \BibitemOpen
  \bibfield  {author} {\bibinfo {author} {\bibfnamefont {C.}~\bibnamefont
  {Joshi}}, \bibinfo {author} {\bibfnamefont {A.}~\bibnamefont {Farsi}},
  \bibinfo {author} {\bibfnamefont {S.}~\bibnamefont {Clemmen}}, \bibinfo
  {author} {\bibfnamefont {S.}~\bibnamefont {Ramelow}}, \ and\ \bibinfo
  {author} {\bibfnamefont {A.~L.}\ \bibnamefont {Gaeta}},\ }\href@noop {}
  {\bibfield  {journal} {\bibinfo  {journal} {Nat. Commun.}\ }\textbf {\bibinfo
  {volume} {9}},\ \bibinfo {pages} {847} (\bibinfo {year} {2018})}\BibitemShut
  {NoStop}%
\bibitem [{\citenamefont {Ansari}\ \emph {et~al.}(2018)\citenamefont {Ansari},
  \citenamefont {Roccia}, \citenamefont {Santandrea}, \citenamefont {Doostdar},
  \citenamefont {Eigner}, \citenamefont {Padberg}, \citenamefont {Gianani},
  \citenamefont {Sbroscia}, \citenamefont {Donohue}, \citenamefont {Mancino}
  \emph {et~al.}}]{Ansari18}%
  \BibitemOpen
  \bibfield  {author} {\bibinfo {author} {\bibfnamefont {V.}~\bibnamefont
  {Ansari}}, \bibinfo {author} {\bibfnamefont {E.}~\bibnamefont {Roccia}},
  \bibinfo {author} {\bibfnamefont {M.}~\bibnamefont {Santandrea}}, \bibinfo
  {author} {\bibfnamefont {M.}~\bibnamefont {Doostdar}}, \bibinfo {author}
  {\bibfnamefont {C.}~\bibnamefont {Eigner}}, \bibinfo {author} {\bibfnamefont
  {L.}~\bibnamefont {Padberg}}, \bibinfo {author} {\bibfnamefont
  {I.}~\bibnamefont {Gianani}}, \bibinfo {author} {\bibfnamefont
  {M.}~\bibnamefont {Sbroscia}}, \bibinfo {author} {\bibfnamefont {J.~M.}\
  \bibnamefont {Donohue}}, \bibinfo {author} {\bibfnamefont {L.}~\bibnamefont
  {Mancino}},  \emph {et~al.},\ }\href@noop {} {\bibfield  {journal} {\bibinfo
  {journal} {Opt. Express}\ }\textbf {\bibinfo {volume} {26}},\ \bibinfo
  {pages} {2764} (\bibinfo {year} {2018})}\BibitemShut {NoStop}%
\bibitem [{\citenamefont {Lee}(1995)}]{Lee95}%
  \BibitemOpen
  \bibfield  {author} {\bibinfo {author} {\bibfnamefont {C.~T.}\ \bibnamefont
  {Lee}},\ }\href {\doibase 10.1103/PhysRevA.52.3374} {\bibfield  {journal}
  {\bibinfo  {journal} {Phys. Rev. A}\ }\textbf {\bibinfo {volume} {52}},\
  \bibinfo {pages} {3374} (\bibinfo {year} {1995})}\BibitemShut {NoStop}%
\bibitem [{\citenamefont {Navarrete-Benlloch}(2015)}]{Navarrete15}%
  \BibitemOpen
  \bibfield  {author} {\bibinfo {author} {\bibfnamefont {C.}~\bibnamefont
  {Navarrete-Benlloch}},\ }\href@noop {} {\emph {\bibinfo {title} {An
  introduction to the formalism of quantum information with continuous
  variables}}}\ (\bibinfo  {publisher} {Morgan and Claypool Publishers},\
  \bibinfo {year} {2015})\BibitemShut {NoStop}%
\bibitem [{\citenamefont {Agarwal}\ and\ \citenamefont
  {Tara}(1992)}]{Agarwal92}%
  \BibitemOpen
  \bibfield  {author} {\bibinfo {author} {\bibfnamefont {G.~S.}\ \bibnamefont
  {Agarwal}}\ and\ \bibinfo {author} {\bibfnamefont {K.}~\bibnamefont {Tara}},\
  }\href {\doibase 10.1103/PhysRevA.46.485} {\bibfield  {journal} {\bibinfo
  {journal} {Phys. Rev. A}\ }\textbf {\bibinfo {volume} {46}},\ \bibinfo
  {pages} {485} (\bibinfo {year} {1992})}\BibitemShut {NoStop}%
\bibitem [{\citenamefont {Kiesel}\ \emph {et~al.}(2008)\citenamefont {Kiesel},
  \citenamefont {Vogel}, \citenamefont {Parigi}, \citenamefont {Zavatta},\ and\
  \citenamefont {Bellini}}]{Kiesel08}%
  \BibitemOpen
  \bibfield  {author} {\bibinfo {author} {\bibfnamefont {T.}~\bibnamefont
  {Kiesel}}, \bibinfo {author} {\bibfnamefont {W.}~\bibnamefont {Vogel}},
  \bibinfo {author} {\bibfnamefont {V.}~\bibnamefont {Parigi}}, \bibinfo
  {author} {\bibfnamefont {A.}~\bibnamefont {Zavatta}}, \ and\ \bibinfo
  {author} {\bibfnamefont {M.}~\bibnamefont {Bellini}},\ }\href@noop {}
  {\bibfield  {journal} {\bibinfo  {journal} {Phys. Rev. A}\ }\textbf {\bibinfo
  {volume} {78}},\ \bibinfo {pages} {021804} (\bibinfo {year}
  {2008})}\BibitemShut {NoStop}%
\bibitem [{\citenamefont {De~Bi\`evre}\ \emph {et~al.}(2019)\citenamefont
  {De~Bi\`evre}, \citenamefont {Horoshko}, \citenamefont {Patera},\ and\
  \citenamefont {Kolobov}}]{Bievre19}%
  \BibitemOpen
  \bibfield  {author} {\bibinfo {author} {\bibfnamefont {S.}~\bibnamefont
  {De~Bi\`evre}}, \bibinfo {author} {\bibfnamefont {D.~B.}\ \bibnamefont
  {Horoshko}}, \bibinfo {author} {\bibfnamefont {G.}~\bibnamefont {Patera}}, \
  and\ \bibinfo {author} {\bibfnamefont {M.~I.}\ \bibnamefont {Kolobov}},\
  }\href {\doibase 10.1103/PhysRevLett.122.080402} {\bibfield  {journal}
  {\bibinfo  {journal} {Phys. Rev. Lett.}\ }\textbf {\bibinfo {volume} {122}},\
  \bibinfo {pages} {080402} (\bibinfo {year} {2019})}\BibitemShut {NoStop}%
\bibitem [{\citenamefont {Horoshko}\ \emph {et~al.}(2019)\citenamefont
  {Horoshko}, \citenamefont {De~Bi\`evre}, \citenamefont {Patera},\ and\
  \citenamefont {Kolobov}}]{Horoshko19WoC}%
  \BibitemOpen
  \bibfield  {author} {\bibinfo {author} {\bibfnamefont {D.}~\bibnamefont
  {Horoshko}}, \bibinfo {author} {\bibfnamefont {S.}~\bibnamefont
  {De~Bi\`evre}}, \bibinfo {author} {\bibfnamefont {G.}~\bibnamefont {Patera}},
  \ and\ \bibinfo {author} {\bibfnamefont {M.}~\bibnamefont {Kolobov}},\ }\href
  {\doibase 10.1051/epjconf/201919800010} {\bibfield  {journal} {\bibinfo
  {journal} {EPJ Web Conf.}\ }\textbf {\bibinfo {volume} {198}},\ \bibinfo
  {pages} {00010} (\bibinfo {year} {2019})}\BibitemShut {NoStop}%
\bibitem [{\citenamefont {Law}\ \emph {et~al.}(2000)\citenamefont {Law},
  \citenamefont {Walmsley},\ and\ \citenamefont {Eberly}}]{Law00}%
  \BibitemOpen
  \bibfield  {author} {\bibinfo {author} {\bibfnamefont {C.~K.}\ \bibnamefont
  {Law}}, \bibinfo {author} {\bibfnamefont {I.~A.}\ \bibnamefont {Walmsley}}, \
  and\ \bibinfo {author} {\bibfnamefont {J.~H.}\ \bibnamefont {Eberly}},\
  }\href {\doibase 10.1103/PhysRevLett.84.5304} {\bibfield  {journal} {\bibinfo
   {journal} {Phys. Rev. Lett.}\ }\textbf {\bibinfo {volume} {84}},\ \bibinfo
  {pages} {5304} (\bibinfo {year} {2000})}\BibitemShut {NoStop}%
\bibitem [{\citenamefont {Horoshko}\ \emph {et~al.}(2012)\citenamefont
  {Horoshko}, \citenamefont {Patera}, \citenamefont {Gatti},\ and\
  \citenamefont {Kolobov}}]{Horoshko12}%
  \BibitemOpen
  \bibfield  {author} {\bibinfo {author} {\bibfnamefont {D.~B.}\ \bibnamefont
  {Horoshko}}, \bibinfo {author} {\bibfnamefont {G.}~\bibnamefont {Patera}},
  \bibinfo {author} {\bibfnamefont {A.}~\bibnamefont {Gatti}}, \ and\ \bibinfo
  {author} {\bibfnamefont {M.~I.}\ \bibnamefont {Kolobov}},\ }\href {\doibase
  10.1140/epjd/e2012-30099-y} {\bibfield  {journal} {\bibinfo  {journal} {Eur.
  Phys. J. D}\ }\textbf {\bibinfo {volume} {66}},\ \bibinfo {pages} {239}
  (\bibinfo {year} {2012})}\BibitemShut {NoStop}%
\bibitem [{\citenamefont {Grice}\ \emph {et~al.}(2001)\citenamefont {Grice},
  \citenamefont {U'Ren},\ and\ \citenamefont {Walmsley}}]{Grice01}%
  \BibitemOpen
  \bibfield  {author} {\bibinfo {author} {\bibfnamefont {W.~P.}\ \bibnamefont
  {Grice}}, \bibinfo {author} {\bibfnamefont {A.~B.}\ \bibnamefont {U'Ren}}, \
  and\ \bibinfo {author} {\bibfnamefont {I.~A.}\ \bibnamefont {Walmsley}},\
  }\href {\doibase 10.1103/PhysRevA.64.063815} {\bibfield  {journal} {\bibinfo
  {journal} {Phys. Rev. A}\ }\textbf {\bibinfo {volume} {64}},\ \bibinfo
  {pages} {063815} (\bibinfo {year} {2001})}\BibitemShut {NoStop}%
\bibitem [{\citenamefont {Barnett}\ and\ \citenamefont
  {Radmore}(1997)}]{BarnettBook}%
  \BibitemOpen
  \bibfield  {author} {\bibinfo {author} {\bibfnamefont {S.~M.}\ \bibnamefont
  {Barnett}}\ and\ \bibinfo {author} {\bibfnamefont {P.~M.}\ \bibnamefont
  {Radmore}},\ }\href@noop {} {\emph {\bibinfo {title} {Methods in theoretical
  quantum optics}}}\ (\bibinfo  {publisher} {Clarendon Press},\ \bibinfo
  {address} {Oxford},\ \bibinfo {year} {1997})\BibitemShut {NoStop}%
\bibitem [{\citenamefont {Titulaer}\ and\ \citenamefont
  {Glauber}(1965)}]{Titulaer65}%
  \BibitemOpen
  \bibfield  {author} {\bibinfo {author} {\bibfnamefont {U.~M.}\ \bibnamefont
  {Titulaer}}\ and\ \bibinfo {author} {\bibfnamefont {R.~J.}\ \bibnamefont
  {Glauber}},\ }\href {\doibase 10.1103/PhysRev.140.B676} {\bibfield  {journal}
  {\bibinfo  {journal} {Phys. Rev.}\ }\textbf {\bibinfo {volume} {140}},\
  \bibinfo {pages} {B676} (\bibinfo {year} {1965})}\BibitemShut {NoStop}%
\bibitem [{\citenamefont {Hogg}\ \emph {et~al.}(2014)\citenamefont {Hogg},
  \citenamefont {Berry},\ and\ \citenamefont {Lvovsky}}]{Hogg14}%
  \BibitemOpen
  \bibfield  {author} {\bibinfo {author} {\bibfnamefont {D.}~\bibnamefont
  {Hogg}}, \bibinfo {author} {\bibfnamefont {D.~W.}\ \bibnamefont {Berry}}, \
  and\ \bibinfo {author} {\bibfnamefont {A.~I.}\ \bibnamefont {Lvovsky}},\
  }\href {\doibase 10.1103/PhysRevA.90.053846} {\bibfield  {journal} {\bibinfo
  {journal} {Phys. Rev. A}\ }\textbf {\bibinfo {volume} {90}},\ \bibinfo
  {pages} {053846} (\bibinfo {year} {2014})}\BibitemShut {NoStop}%
\bibitem [{\citenamefont {Laurat}\ \emph {et~al.}(2004)\citenamefont {Laurat},
  \citenamefont {Coudreau}, \citenamefont {Treps}, \citenamefont
  {Ma\^{\i}tre},\ and\ \citenamefont {Fabre}}]{Laurat04}%
  \BibitemOpen
  \bibfield  {author} {\bibinfo {author} {\bibfnamefont {J.}~\bibnamefont
  {Laurat}}, \bibinfo {author} {\bibfnamefont {T.}~\bibnamefont {Coudreau}},
  \bibinfo {author} {\bibfnamefont {N.}~\bibnamefont {Treps}}, \bibinfo
  {author} {\bibfnamefont {A.}~\bibnamefont {Ma\^{\i}tre}}, \ and\ \bibinfo
  {author} {\bibfnamefont {C.}~\bibnamefont {Fabre}},\ }\href {\doibase
  10.1103/PhysRevA.69.033808} {\bibfield  {journal} {\bibinfo  {journal} {Phys.
  Rev. A}\ }\textbf {\bibinfo {volume} {69}},\ \bibinfo {pages} {033808}
  (\bibinfo {year} {2004})}\BibitemShut {NoStop}%
\bibitem [{\citenamefont {D'Auria}\ \emph {et~al.}(2012)\citenamefont
  {D'Auria}, \citenamefont {Morin}, \citenamefont {Fabre},\ and\ \citenamefont
  {Laurat}}]{DAuria12}%
  \BibitemOpen
  \bibfield  {author} {\bibinfo {author} {\bibfnamefont {V.}~\bibnamefont
  {D'Auria}}, \bibinfo {author} {\bibfnamefont {O.}~\bibnamefont {Morin}},
  \bibinfo {author} {\bibfnamefont {C.}~\bibnamefont {Fabre}}, \ and\ \bibinfo
  {author} {\bibfnamefont {J.}~\bibnamefont {Laurat}},\ }\href {\doibase
  10.1140/epjd/e2012-30351-6} {\bibfield  {journal} {\bibinfo  {journal} {Eur.
  Phys. J. D}\ }\textbf {\bibinfo {volume} {66}},\ \bibinfo {pages} {249}
  (\bibinfo {year} {2012})}\BibitemShut {NoStop}%
\bibitem [{\citenamefont {Quesada}(2015)}]{Quesada15Thesis}%
  \BibitemOpen
  \bibfield  {author} {\bibinfo {author} {\bibfnamefont {N.}~\bibnamefont
  {Quesada}},\ }\emph {\bibinfo {title} {Very nonlinear quantum optics}},\
  \href@noop {} {Ph.D. thesis},\ \bibinfo  {school} {University of Toronto}
  (\bibinfo {year} {2015})\BibitemShut {NoStop}%
\bibitem [{\citenamefont {Tiedau}\ \emph {et~al.}(2019)\citenamefont {Tiedau},
  \citenamefont {Bartley}, \citenamefont {Harder}, \citenamefont {Lita},
  \citenamefont {Nam}, \citenamefont {Gerrits},\ and\ \citenamefont
  {Silberhorn}}]{Tiedau19}%
  \BibitemOpen
  \bibfield  {author} {\bibinfo {author} {\bibfnamefont {J.}~\bibnamefont
  {Tiedau}}, \bibinfo {author} {\bibfnamefont {T.~J.}\ \bibnamefont {Bartley}},
  \bibinfo {author} {\bibfnamefont {G.}~\bibnamefont {Harder}}, \bibinfo
  {author} {\bibfnamefont {A.~E.}\ \bibnamefont {Lita}}, \bibinfo {author}
  {\bibfnamefont {S.~W.}\ \bibnamefont {Nam}}, \bibinfo {author} {\bibfnamefont
  {T.}~\bibnamefont {Gerrits}}, \ and\ \bibinfo {author} {\bibfnamefont
  {C.}~\bibnamefont {Silberhorn}},\ }\href@noop {} {\bibfield  {journal}
  {\bibinfo  {journal} {arXiv preprint arXiv:1901.03237}\ } (\bibinfo {year}
  {2019})}\BibitemShut {NoStop}%
\bibitem [{\citenamefont {Zavatta}\ \emph {et~al.}(2008)\citenamefont
  {Zavatta}, \citenamefont {Parigi}, \citenamefont {Kim},\ and\ \citenamefont
  {Bellini}}]{Zavatta08}%
  \BibitemOpen
  \bibfield  {author} {\bibinfo {author} {\bibfnamefont {A.}~\bibnamefont
  {Zavatta}}, \bibinfo {author} {\bibfnamefont {V.}~\bibnamefont {Parigi}},
  \bibinfo {author} {\bibfnamefont {M.~S.}\ \bibnamefont {Kim}}, \ and\
  \bibinfo {author} {\bibfnamefont {M.}~\bibnamefont {Bellini}},\ }\href
  {\doibase 10.1088/1367-2630/10/12/123006} {\bibfield  {journal} {\bibinfo
  {journal} {New J. Phys.}\ }\textbf {\bibinfo {volume} {10}},\ \bibinfo
  {pages} {123006} (\bibinfo {year} {2008})}\BibitemShut {NoStop}%
\bibitem [{\citenamefont {Cahill}\ and\ \citenamefont
  {Glauber}(1969)}]{Cahill69b}%
  \BibitemOpen
  \bibfield  {author} {\bibinfo {author} {\bibfnamefont {K.~E.}\ \bibnamefont
  {Cahill}}\ and\ \bibinfo {author} {\bibfnamefont {R.~J.}\ \bibnamefont
  {Glauber}},\ }\href {\doibase 10.1103/PhysRev.177.1882} {\bibfield  {journal}
  {\bibinfo  {journal} {Phys. Rev.}\ }\textbf {\bibinfo {volume} {177}},\
  \bibinfo {pages} {1882} (\bibinfo {year} {1969})}\BibitemShut {NoStop}%
\bibitem [{\citenamefont {Damanet}\ \emph {et~al.}(2018)\citenamefont
  {Damanet}, \citenamefont {K\"ubler}, \citenamefont {Martin},\ and\
  \citenamefont {Braun}}]{Damanet18}%
  \BibitemOpen
  \bibfield  {author} {\bibinfo {author} {\bibfnamefont {F.}~\bibnamefont
  {Damanet}}, \bibinfo {author} {\bibfnamefont {J.}~\bibnamefont {K\"ubler}},
  \bibinfo {author} {\bibfnamefont {J.}~\bibnamefont {Martin}}, \ and\ \bibinfo
  {author} {\bibfnamefont {D.}~\bibnamefont {Braun}},\ }\href {\doibase
  10.1103/PhysRevA.97.023832} {\bibfield  {journal} {\bibinfo  {journal} {Phys.
  Rev. A}\ }\textbf {\bibinfo {volume} {97}},\ \bibinfo {pages} {023832}
  (\bibinfo {year} {2018})}\BibitemShut {NoStop}%
\bibitem [{\citenamefont {Kolobov}(1999)}]{Kolobov99}%
  \BibitemOpen
  \bibfield  {author} {\bibinfo {author} {\bibfnamefont {M.~I.}\ \bibnamefont
  {Kolobov}},\ }\href {\doibase 10.1103/RevModPhys.71.1539} {\bibfield
  {journal} {\bibinfo  {journal} {Rev. Mod. Phys.}\ }\textbf {\bibinfo {volume}
  {71}},\ \bibinfo {pages} {1539} (\bibinfo {year} {1999})}\BibitemShut
  {NoStop}%
\bibitem [{\citenamefont {Yurke}\ and\ \citenamefont {Stoler}(1986)}]{Yurke86}%
  \BibitemOpen
  \bibfield  {author} {\bibinfo {author} {\bibfnamefont {B.}~\bibnamefont
  {Yurke}}\ and\ \bibinfo {author} {\bibfnamefont {D.}~\bibnamefont {Stoler}},\
  }\href {\doibase 10.1103/PhysRevLett.57.13} {\bibfield  {journal} {\bibinfo
  {journal} {Phys. Rev. Lett.}\ }\textbf {\bibinfo {volume} {57}},\ \bibinfo
  {pages} {13} (\bibinfo {year} {1986})}\BibitemShut {NoStop}%
\bibitem [{\citenamefont {Horoshko}\ and\ \citenamefont
  {Kilin}(1998)}]{Horoshko98}%
  \BibitemOpen
  \bibfield  {author} {\bibinfo {author} {\bibfnamefont {D.~B.}\ \bibnamefont
  {Horoshko}}\ and\ \bibinfo {author} {\bibfnamefont {S.~Y.}\ \bibnamefont
  {Kilin}},\ }\href {\doibase 10.1364/OE.2.000347} {\bibfield  {journal}
  {\bibinfo  {journal} {Opt. Express}\ }\textbf {\bibinfo {volume} {2}},\
  \bibinfo {pages} {347} (\bibinfo {year} {1998})}\BibitemShut {NoStop}%
\bibitem [{\citenamefont {Haroche}\ and\ \citenamefont
  {Raimond}(2006)}]{HarocheBook}%
  \BibitemOpen
  \bibfield  {author} {\bibinfo {author} {\bibfnamefont {S.}~\bibnamefont
  {Haroche}}\ and\ \bibinfo {author} {\bibfnamefont {J.-M.}\ \bibnamefont
  {Raimond}},\ }\href@noop {} {\emph {\bibinfo {title} {Exploring the Quantum:
  Atoms, Cavities and Photons}}}\ (\bibinfo  {publisher} {Oxford University
  Press},\ \bibinfo {year} {2006})\BibitemShut {NoStop}%
\bibitem [{\citenamefont {Horoshko}\ \emph {et~al.}(2016)\citenamefont
  {Horoshko}, \citenamefont {De~Bi\`evre}, \citenamefont {Kolobov},\ and\
  \citenamefont {Patera}}]{Horoshko16}%
  \BibitemOpen
  \bibfield  {author} {\bibinfo {author} {\bibfnamefont {D.~B.}\ \bibnamefont
  {Horoshko}}, \bibinfo {author} {\bibfnamefont {S.}~\bibnamefont
  {De~Bi\`evre}}, \bibinfo {author} {\bibfnamefont {M.~I.}\ \bibnamefont
  {Kolobov}}, \ and\ \bibinfo {author} {\bibfnamefont {G.}~\bibnamefont
  {Patera}},\ }\href {\doibase 10.1103/PhysRevA.93.062323} {\bibfield
  {journal} {\bibinfo  {journal} {Phys. Rev. A}\ }\textbf {\bibinfo {volume}
  {93}},\ \bibinfo {pages} {062323} (\bibinfo {year} {2016})}\BibitemShut
  {NoStop}%
\bibitem [{\citenamefont {Kim}\ \emph {et~al.}(1989)\citenamefont {Kim},
  \citenamefont {de~Oliveira},\ and\ \citenamefont {Knight}}]{Kim89}%
  \BibitemOpen
  \bibfield  {author} {\bibinfo {author} {\bibfnamefont {M.~S.}\ \bibnamefont
  {Kim}}, \bibinfo {author} {\bibfnamefont {F.~A.~M.}\ \bibnamefont
  {de~Oliveira}}, \ and\ \bibinfo {author} {\bibfnamefont {P.~L.}\ \bibnamefont
  {Knight}},\ }\href {\doibase 10.1103/PhysRevA.40.2494} {\bibfield  {journal}
  {\bibinfo  {journal} {Phys. Rev. A}\ }\textbf {\bibinfo {volume} {40}},\
  \bibinfo {pages} {2494} (\bibinfo {year} {1989})}\BibitemShut {NoStop}%
\bibitem [{\citenamefont {Lee}\ and\ \citenamefont {Jeong}(2011)}]{Lee11}%
  \BibitemOpen
  \bibfield  {author} {\bibinfo {author} {\bibfnamefont {C.-W.}\ \bibnamefont
  {Lee}}\ and\ \bibinfo {author} {\bibfnamefont {H.}~\bibnamefont {Jeong}},\
  }\href {\doibase 10.1103/PhysRevLett.106.220401} {\bibfield  {journal}
  {\bibinfo  {journal} {Phys. Rev. Lett.}\ }\textbf {\bibinfo {volume} {106}},\
  \bibinfo {pages} {220401} (\bibinfo {year} {2011})}\BibitemShut {NoStop}%
\bibitem [{\citenamefont {Hillery}(1989)}]{Hillery89}%
  \BibitemOpen
  \bibfield  {author} {\bibinfo {author} {\bibfnamefont {M.}~\bibnamefont
  {Hillery}},\ }\href {\doibase 10.1103/PhysRevA.39.2994} {\bibfield  {journal}
  {\bibinfo  {journal} {Phys. Rev. A}\ }\textbf {\bibinfo {volume} {39}},\
  \bibinfo {pages} {2994} (\bibinfo {year} {1989})}\BibitemShut {NoStop}%
\bibitem [{\citenamefont {Quesada}\ \emph {et~al.}(2019)\citenamefont
  {Quesada}, \citenamefont {Helt}, \citenamefont {Izaac}, \citenamefont
  {Arrazola}, \citenamefont {Shahrokhshahi}, \citenamefont {Myers},\ and\
  \citenamefont {Sabapathy}}]{Quesada19}%
  \BibitemOpen
  \bibfield  {author} {\bibinfo {author} {\bibfnamefont {N.}~\bibnamefont
  {Quesada}}, \bibinfo {author} {\bibfnamefont {L.~G.}\ \bibnamefont {Helt}},
  \bibinfo {author} {\bibfnamefont {J.}~\bibnamefont {Izaac}}, \bibinfo
  {author} {\bibfnamefont {J.~M.}\ \bibnamefont {Arrazola}}, \bibinfo {author}
  {\bibfnamefont {R.}~\bibnamefont {Shahrokhshahi}}, \bibinfo {author}
  {\bibfnamefont {C.~R.}\ \bibnamefont {Myers}}, \ and\ \bibinfo {author}
  {\bibfnamefont {K.~K.}\ \bibnamefont {Sabapathy}},\ }\href {\doibase
  10.1103/PhysRevA.100.022341} {\bibfield  {journal} {\bibinfo  {journal}
  {Phys. Rev. A}\ }\textbf {\bibinfo {volume} {100}},\ \bibinfo {pages}
  {022341} (\bibinfo {year} {2019})}\BibitemShut {NoStop}%
\bibitem [{\citenamefont {Yadin}\ \emph {et~al.}(2018)\citenamefont {Yadin},
  \citenamefont {Binder}, \citenamefont {Thompson}, \citenamefont
  {Narasimhachar}, \citenamefont {Gu},\ and\ \citenamefont {Kim}}]{Yadin18}%
  \BibitemOpen
  \bibfield  {author} {\bibinfo {author} {\bibfnamefont {B.}~\bibnamefont
  {Yadin}}, \bibinfo {author} {\bibfnamefont {F.~C.}\ \bibnamefont {Binder}},
  \bibinfo {author} {\bibfnamefont {J.}~\bibnamefont {Thompson}}, \bibinfo
  {author} {\bibfnamefont {V.}~\bibnamefont {Narasimhachar}}, \bibinfo {author}
  {\bibfnamefont {M.}~\bibnamefont {Gu}}, \ and\ \bibinfo {author}
  {\bibfnamefont {M.~S.}\ \bibnamefont {Kim}},\ }\href {\doibase
  10.1103/PhysRevX.8.041038} {\bibfield  {journal} {\bibinfo  {journal} {Phys.
  Rev. X}\ }\textbf {\bibinfo {volume} {8}},\ \bibinfo {pages} {041038}
  (\bibinfo {year} {2018})}\BibitemShut {NoStop}%
\bibitem [{\citenamefont {Kwon}\ \emph {et~al.}(2019)\citenamefont {Kwon},
  \citenamefont {Tan}, \citenamefont {Volkoff},\ and\ \citenamefont
  {Jeong}}]{Kwon19}%
  \BibitemOpen
  \bibfield  {author} {\bibinfo {author} {\bibfnamefont {H.}~\bibnamefont
  {Kwon}}, \bibinfo {author} {\bibfnamefont {K.~C.}\ \bibnamefont {Tan}},
  \bibinfo {author} {\bibfnamefont {T.}~\bibnamefont {Volkoff}}, \ and\
  \bibinfo {author} {\bibfnamefont {H.}~\bibnamefont {Jeong}},\ }\href
  {\doibase 10.1103/PhysRevLett.122.040503} {\bibfield  {journal} {\bibinfo
  {journal} {Phys. Rev. Lett.}\ }\textbf {\bibinfo {volume} {122}},\ \bibinfo
  {pages} {040503} (\bibinfo {year} {2019})}\BibitemShut {NoStop}%
\bibitem [{\citenamefont {Kenfack}\ and\ \citenamefont
  {\.{Z}yczkowski}(2004)}]{Kenfack04}%
  \BibitemOpen
  \bibfield  {author} {\bibinfo {author} {\bibfnamefont {A.}~\bibnamefont
  {Kenfack}}\ and\ \bibinfo {author} {\bibfnamefont {K.}~\bibnamefont
  {\.{Z}yczkowski}},\ }\href {\doibase 10.1088/1464-4266/6/10/003} {\bibfield
  {journal} {\bibinfo  {journal} {J. Opt. B}\ }\textbf {\bibinfo {volume}
  {6}},\ \bibinfo {pages} {396} (\bibinfo {year} {2004})}\BibitemShut {NoStop}%
\end{thebibliography}%

\end{document}